\documentclass[aip,jcp,reprint,noshowkeys,superscriptaddress]{revtex4-2}
\usepackage{graphicx,dcolumn,bm,microtype,multirow,amscd,amsmath,amssymb,amsfonts,physics,wrapfig,bbold,siunitx,xspace,txfonts}
\usepackage[version=4]{mhchem}

\usepackage[utf8]{inputenc}
\usepackage[T1]{fontenc}
\usepackage{feynmp-auto}
\usepackage{longtable}
\usepackage{threeparttable}
\usepackage[dvipsnames]{xcolor}
\allowdisplaybreaks[4]
\newlength{\ipnumberwidth}
\newcommand{\statcell}[1]{
  \multicolumn{1}{c}{
    \settowidth{\ipnumberwidth}{$0000.00$}
    \makebox[\ipnumberwidth][r]{$#1$}
  }
}

\usepackage{hyperref}
\hypersetup{
    colorlinks,
    linkcolor={red!50!black},
    citecolor={red!70!black},
    urlcolor={red!80!black}
}

\usepackage{listings}
\definecolor{codegreen}{rgb}{0.58,0.4,0.2}
\definecolor{codegray}{rgb}{0.5,0.5,0.5}
\definecolor{codepurple}{rgb}{0.25,0.35,0.55}
\definecolor{codeblue}{rgb}{0.30,0.60,0.8}
\definecolor{backcolour}{rgb}{0.98,0.98,0.98}
\definecolor{mygray}{rgb}{0.5,0.5,0.5}

\definecolor{sqred}{rgb}{0.85,0.1,0.1}
\definecolor{sqgreen}{rgb}{0.25,0.65,0.15}
\definecolor{sqorange}{rgb}{0.90,0.50,0.15}
\definecolor{sqblue}{rgb}{0.10,0.3,0.60}

\lstdefinestyle{mystyle}{
    backgroundcolor=\color{backcolour},
    commentstyle=\color{codegreen},
    keywordstyle=\color{codeblue},
    numberstyle=\tiny\color{codegray},
    stringstyle=\color{codepurple},
    basicstyle=\ttfamily\footnotesize,
    breakatwhitespace=false,
    breaklines=true,
    captionpos=b,
    keepspaces=true,
    numbers=left,
    numbersep=5pt,
    numberstyle=\ttfamily\tiny\color{mygray},
    showspaces=false,
    showstringspaces=false,
    showtabs=false,
    tabsize=2
  }

  \newcolumntype{d}{D{.}{.}{-1}}

\newcommand{\cre}[1]{\hat{a}_{#1}^\dagger}
\newcommand{\ani}[1]{\hat{a}_{#1}}

\newcommand{\NO}[1]{\{#1\}} 
\newcommand{\ERI}[2]{v^{#1}_{#2}}
\newcommand{\aERI}[2]{\bar{v}^{#1}_{#2}}
\newcommand{\sERI}[2]{(#1|#2)}

\newcommand{\fnt}{\footnotetext}
\newcommand{\fnm}{\footnotemark}

\newcommand{\br}{\boldsymbol{r}}
\newcommand{\bx}{\boldsymbol{x}}

\newcommand{\bA}{\boldsymbol{A}}
\newcommand{\bB}{\boldsymbol{B}}
\newcommand{\bC}{\boldsymbol{C}}
\newcommand{\bK}{\boldsymbol{K}}
\newcommand{\bX}{\boldsymbol{X}}
\newcommand{\bY}{\boldsymbol{Y}}
\newcommand{\bT}{\boldsymbol{T}}

\newcommand{\bH}{\boldsymbol{H}}
\newcommand{\bO}{\boldsymbol{0}}
\newcommand{\bI}{\boldsymbol{1}}
\newcommand{\bLam}{\boldsymbol{\Lambda}}

\newcommand{\bOm}{\boldsymbol{\Omega}}

\newcommand{\hT}{\Hat{T}}

\newcommand{\hR}{\Hat{R}}
\newcommand{\hH}{\Hat{H}}

\newcommand{\hLam}{\Hat{\Lambda}}

\newcommand{\wtH}{\widetilde{H}}

\newcommand{\wWd}{W}
\newcommand{\wW}{\widetilde{W}}
\newcommand{\wF}{\tilde{f}}
\newcommand{\wFF}{\widetilde{F}}
\newcommand{\wVd}{V}

\newcommand{\xc}{\text{xc}}

\newcommand{\ldr}{\text{$\lambda$-dr}}

\newcommand{\ii}{\text{i}}

\makeatletter
\newcommand{\tablecaption}[1]{
	\begingroup
	\def\@captype{table}
	\caption{#1}
	\endgroup
}
\makeatother

\newcommand{\LCPQ}{Univ Toulouse, CNRS, Laboratoire de Chimie et Physique Quantiques, Toulouse, France}

\begin{document}	

\title{Bringing Singles to $GW$: A Coupled-Cluster Perspective on Orbital Relaxation}

\author{Marios-Petros \surname{Kitsaras}}
	\email{kitsaras@irsamc.ups-tlse.fr}
	\affiliation{\LCPQ}

\author{Loris \surname{Burth}}
		\email{lburth@irsamc.ups-tlse.fr}
		\affiliation{\LCPQ}
	
\author{Pierre-Fran\c{c}ois \surname{Loos}}
	\email{loos@irsamc.ups-tlse.fr}
	\affiliation{\LCPQ}

\begin{abstract}
The one-shot $G_0W_0$ approximation suffers from a pronounced dependence on the underlying mean-field starting point. 
Here, we explore an alternative to conventional self-consistent $GW$ schemes by introducing single excitations within the recently established connection between $GW$ and extended coupled-cluster (ECC) theory. 
Starting from the equivalence between $G_0W_0$ and ionization-potential/electron-affinity equation-of-motion direct-ring extended coupled-cluster doubles (IP/EA-EOM-drECCD), 
we extend the formalism to include right-hand single-excitation and left-hand single-deexcitation cluster amplitudes coupled to the direct-ring doubles amplitudes.
The resulting approximation, denoted singles-$G_0W_0$ (s$G_0W_0$), incorporates orbital-relaxation effects without explicitly introducing self-consistency in the Green's function.
We assess the performance of s$G_0W_0$ for molecular valence and core ionization potentials using Hartree--Fock and Kohn--Sham starting points. 
For both types of ionizations, the inclusion of singles largely suppresses the starting-point dependence observed in conventional $G_0W_0$, yielding quasiparticle energies close to those obtained from the Hartree--Fock reference, irrespective of the initial mean-field description. 
Although this does not necessarily translate into improved absolute accuracy, it demonstrates that singles provide an effective route to reducing the starting-point dependence of $GW$-like approximations without resorting to self-consistency.
 \bigskip
 \begin{center}
 	\boxed{\includegraphics[width=0.5\linewidth]{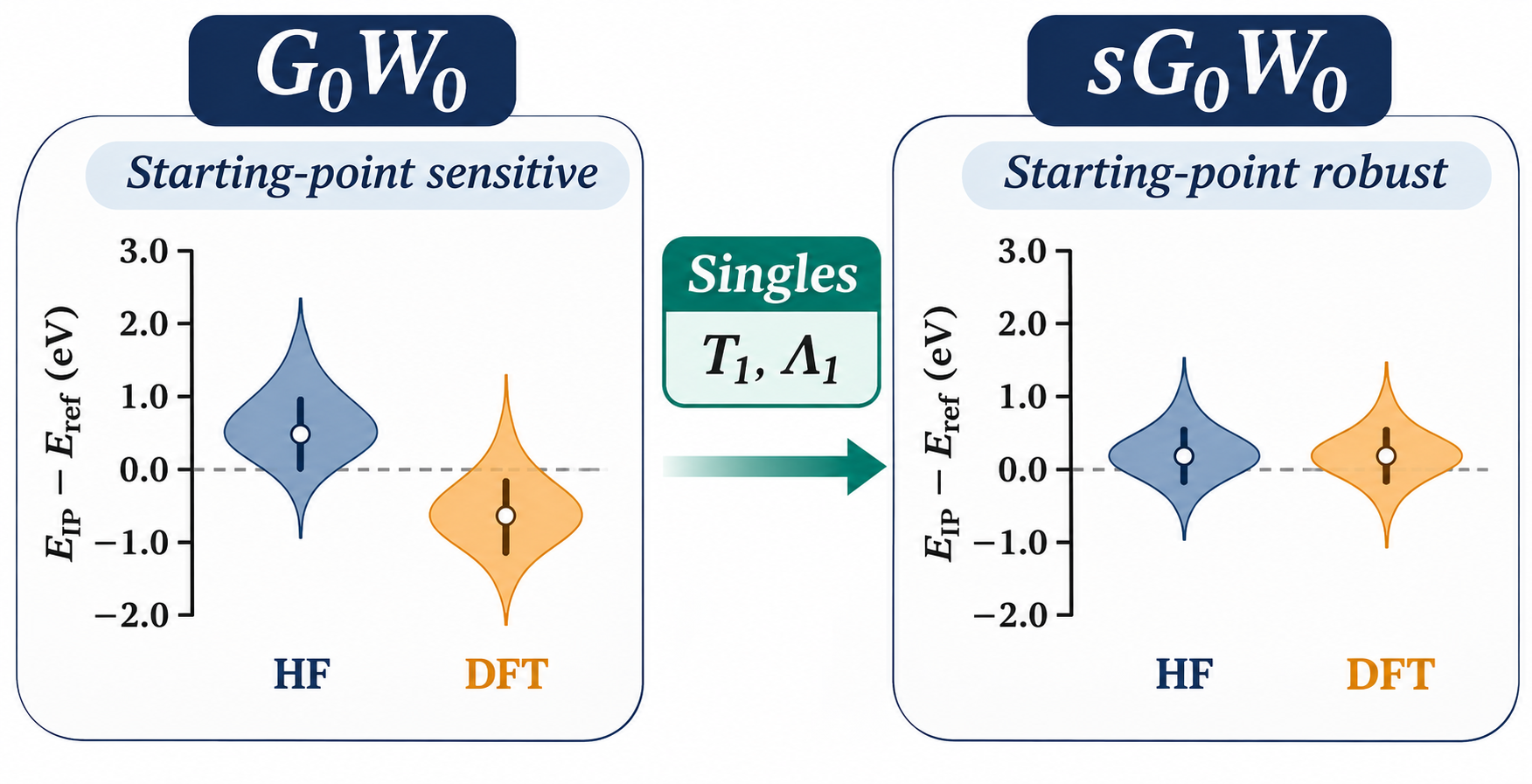}}
 \end{center}
 \bigskip
\end{abstract}

\maketitle

\section{Introduction}

The $GW$ approximation has become one of the most widely used approaches to the calculation of charged electronic excitations in molecules and extended systems.\cite{Hedin1965,Aryasetiawan1998,Onida2002,Reining2018,MartinReiningCeperley2016,Golze2019,MarieAmmarLoos2024,Hedin1999}
Originally introduced by Hedin as an approximation to the electronic self-energy, it describes the propagation of an electron or hole in terms of the one-body Green's function $G$ and the dynamically screened Coulomb interaction $W$.\cite{Hedin1965,HedinLundqvist1969,StrinatiMattauschHanke1980,HybertsenLouie1985}
Screening is usually evaluated within the random-phase approximation (RPA), \cite{Bohm_1951,Bohm_1953,Nozieres_1958,Pines_1952,RenRinkeJoas2012,Chen_2017} which resums an infinite series of polarization diagrams --- the so-called ring or bubble diagrams --- and thereby captures an essential part of the electronic response in the weakly-correlated regime.\cite{MattuckBook,Gell-Mann_1957}
Efficient stochastic, low-scaling, and auxiliary-space algorithms have further broadened the range of systems accessible to $GW$.\cite{NeuhauserEtAl2014,GovoniGalli2015,VlcekEtAl2017,WilhelmEtAl2018,ForsterVisscher2020,DucheminBlase2021,ToelleNiemeyerNeugebauer2024}
The combination of $GW$ quasiparticle energies with the Bethe-Salpeter equation also provides an established framework for optical excitations.\cite{Strinati_1988,Onida2002,MartinReiningCeperley2016,BlaseDucheminJacquemin2018,BlaseDucheminJacqueminLoos2020}
The resulting compromise between physical content and computational cost has made $GW$ particularly attractive for ionization potentials (IPs), electron affinities (EAs), quasiparticle band structures, and fundamental gaps.\cite{HybertsenLouie1986,Godby1988,FaberEtAl2011,Bruneval2012,MaromEtAl2012,vanSettenEtAl2015,CarusoEtAl2016,RangelEtAl2016,BrunevalMarques2013,Golze2019,MoninoLoos2023,OrlandoRomanielloLoos2023,LoosToelle2026}

Despite these successes, $GW$ remains an approximation.
In particular, the neglect of vertex corrections is responsible for deficiencies such as self-screening and an incomplete description of multiparticle features in the photoemission spectrum.\cite{HolmAryasetiawan1997,DelSoleReiningGodby1994,SchindlmayrGodby1998,RomanielloGuyotReining2009,RomanielloBechstedtReining2012,RenMaromCarusoSchefflerRinke2015,GruneisEtAl2014,MaggioKresse2017,LewisBerkelbach2019,LoosRomanielloBerger2018,marie2024val,LoosMarieAmmar2024Cumulant,AmmarEtAl2024,BrunevalForster2024,ForsterBruneval2024,BrunevalForsterPavlyukh2025,Forster2025Vertex,MarieToelleLoos2026,WangEtAl2026}
Moreover, although Hedin's equations suggest a self-consistent formulation, most practical applications rely on the one-shot $G_0W_0$ approximation.
In this scheme, both $G_0$ and $W_0$ are constructed from a fixed mean-field reference, most commonly Hartree--Fock (HF) or Kohn--Sham (KS) density-functional theory.
As a consequence, the calculated quasiparticle energies can depend strongly on the starting-point orbitals and orbital energies.\cite{MaromEtAl2012,BrunevalMarques2013,vanSettenEtAl2015,KnightEtAl2016,CarusoEtAl2016,Golze2019,MarieAmmarLoos2024}

This starting-point dependence can, to some extent, be turned into an advantage.
A suitable KS reference, often obtained from a hybrid functional, can yield significantly more accurate $G_0W_0$ quasiparticle energies than a semilocal KS or HF reference.\cite{MaromEtAl2012,KoerzdoerferMarom2012,BrunevalMarques2013,GallandiEtAl2016,KnightEtAl2016,McKeonEtAl2022}
In particular, changing the fraction of exact exchange modifies the starting-point gap and therefore both the poles of $G_0$ and the amount of screening entering $W_0$.
Global and range-separated hybrids, including nonempirically tuned functionals, have consequently been used to improve the description of molecular IPs and EAs.\cite{GallandiEtAl2016,KnightEtAl2016,BoisKoerzdoerfer2017,MejiaRodriguez2024,SinghJanaSmiga2026,DauthEtAl2016}
Such choices preserve the simplicity of a one-shot calculation but leave open the question of which starting point is best suited to a given system.
However, no single starting point can be expected to provide uniformly accurate results across different systems and electronic states.

A more systematic way to reduce this dependence is to introduce self-consistency.
In eigenvalue-self-consistent $GW$ (ev$GW$), the energies entering the Green's function and screened interaction are updated while the orbitals are kept fixed.
This often reduces the sensitivity to the initial spectrum, although the dependence on the reference orbitals remains.\cite{Golze2019,KnightEtAl2016,CarusoEtAl2016,LiJinRinkeYangGolze2022,LoosRomanielloBerger2018,MoninoLoos2022Discontinuities}
Quasiparticle-self-consistent $GW$ (qs$GW$) additionally updates the orbitals through an effective static, Hermitian approximation to the $GW$ self-energy.\cite{FaleevVanSchilfgaardeKotani2004,vanSchilfgaarde2006,Kotani2007,KaplanEtAl2016,ForsterVisscher2021,MarieLoos2023SRG,LoosToelle2026}
This eliminates starting-point dependence, provided that the iterations converge to the same self-consistent solution.
However, the Green's function retains an independent-particle form with unit quasiparticle weights and no satellite contributions.
Fully self-consistent $GW$ (sc$GW$) removes this last restriction by iterating Dyson's equation for the interacting $G$ and updating $W$ accordingly.\cite{CarusoRinkeRenRubioScheffler2012,BrunevalVastReining2006,ShishkinKresse2007,StanDahlen2006,Rostgaard2010,Caruso2013,CarusoEtAl2016,YehEtAl2022,HarshaEtAl2024,WenEtAl2024,WenHarshaZgid2026}
This treatment is formally independent of the initial reference once a particular self-consistent solution has been reached, but it is substantially more demanding.
The frequency dependence of the Green's function must be retained, and many implementations exploit imaginary-time and Matsubara-frequency representations, frequently at finite temperature.\cite{YehEtAl2022,AbrahamHarshaZgid2024,PokhilkoIskakovYehZgid2021,PokhilkoEtAl2024THC,IskakovEtAl2025,PokhilkoEtAl2025Vertex}
The nonlinear equations may also display slow convergence or multiple solutions, even for relatively small systems.\cite{PokhilkoZgid2021,PokhilkoYehZgid2022,PokhilkoZgid2025}
Importantly, self-consistency alone does not restore the missing vertex corrections, which has motivated many approaches that improve the screening or the self-energy beyond $GW$.\cite{DelSoleReiningGodby1994,SchindlmayrGodby1998,RomanielloGuyotReining2009,GruneisEtAl2014,MaggioKresse2017,WangRinkeRen2021,BrunevalForster2024,ForsterBruneval2024,Forster2025Vertex}
Its effect on accuracy is therefore property- and system-dependent: improvements have been documented in rigorous molecular comparisons, notably in recent work by Zgid and coworkers,\cite{CarusoEtAl2016,HarshaEtAl2024,WenEtAl2024,WenCorrection2025,ForsterBruneval2024} whereas other benchmarks and model studies have identified cases where a carefully chosen one-shot or partially self-consistent calculation is competitive or superior.\cite{KnightEtAl2016,DauthEtAl2016,LoosRomanielloBerger2018,WangEtAl2026}
Thus, removing the reference dependence is not, by itself, a guarantee of better quasiparticle energies.
Nevertheless, reducing this dependence is valuable in its own right, as it limits the arbitrariness associated with the choice of mean-field reference.

Another possibility to reduce starting-point dependence is to account for single-excitation effects without carrying out self-consistency.
The importance of singles was already recognized in the context of RPA correlation energies, where single-excitation and renormalized-singles corrections can markedly improve results obtained from KS orbitals.\cite{RenTkatchenko2011,RenRinkeJoas2012,RenRinkeScuseria2013,PaierEtAl2012}
Building on related ideas, Jin, Su, and Yang introduced the renormalized-singles (RS) Green's function for quasiparticle calculations.\cite{JinSuYang2019}
In this approach, an HF-like Hamiltonian constructed from the KS density matrix is diagonalized separately within the occupied and virtual subspaces.
This procedure changes the orbital energies and allows rotations within each subspace, but it preserves the occupied-virtual partition and hence the underlying reference determinant.
The resulting $G_{\text{RS}}W_0$ and $G_{\text{RS}}W_{\text{RS}}$ schemes can substantially reduce the dependence on the initial KS functional.
The RS construction has also been extended to the $T$-matrix self-energy, the Bethe--Salpeter equation, and renormalized-singles approaches including correlation effects.\cite{LiChenYang2021,LiYang2022,LiGolzeYang2022,LiJinRinkeYangGolze2022}
These developments demonstrate that the inclusion of singles effects provides a promising route to reducing the starting-point dependence of $GW$ without resorting to self-consistency.

A complementary perspective on the role of single excitations is provided by coupled-cluster (CC) theory.\cite{PurvisBartlett1982,BartlettMusial2007,Arponen1983,ArponenBishopPajanne1987I,ArponenBishopPajanne1987II}
Although CC and Green's function many-body perturbation theory have traditionally developed along different lines, their diagrammatic content is closely related.
Scuseria and coworkers showed that the direct-ring coupled-cluster doubles (drCCD) equations reproduce the RPA correlation energy.\cite{Freeman1977,ScuseriaHendersonSorensen2008,JansenLiuAngyan2010,ScuseriaHendersonBulik2013,PengSteinmannVanAggelenYang2013,PaierEtAl2012}
Lange and Berkelbach subsequently compared the $GW$ and IP/EA equation-of-motion CC (EOM-CC) self-energies, highlighting the interplay between ring diagrams and vertex corrections.\cite{Berkelbach2018RPA,RishiPereraBartlett2020,LangeBerkelbach2018}
Further connections have been established through the supermatrix representation of $GW$, the quasiboson formalism, and CC-like formulations of many-body perturbation theory.\cite{BintrimBerkelbach2021,QuinteroMoninoMarieLoos2022,ToelleChan2023,CoveneyTew2025,Coveney2025SelfEnergy,CoveneyTew2026,MarieLoos2025Parquet}
Most recently, some of us demonstrated that, in the appropriate electron-boson representation, the $G_0W_0$ quasiparticle spectrum can be recovered exactly using IP/EA-EOM direct-ring extended coupled-cluster doubles (drECCD).\cite{Toelle2025GWGradients,ToelleKitsarasLoos2025BSEGradients,KitsarasToelleLoos2026,ToelleEtAl2026}
This correspondence offers a well-defined starting point for incorporating additional CC ingredients into a $GW$-like construction.

An important ingredient of conventional CC theory is the single-excitation operator.
Single-reference CC calculations are most commonly performed using HF orbitals, for which Brillouin's theorem makes the reference a natural starting point.
However, Brillouin's theorem does not imply that the correlated energy is stationary with respect to occupied-virtual orbital rotations.
In CCSD, these effects can be accommodated through the $\hT_1$ operator, whose amplitudes are determined simultaneously with the doubles amplitudes and thus respond to electron correlation.\cite{PurvisBartlett1982,BartlettMusial2007}
By Thouless' theorem, $e^{\hT_1}$ acting on a Slater determinant generates another nonorthogonal determinant, providing an orbital-relaxation interpretation of the singles operator.\cite{Thouless1960}
Early orbital-optimization studies showed that the presence of $e^{\hT_1}$ accounts for much of the orbital-relaxation contribution in CCSD.\cite{ScuseriaSchaefer1987}
The explicit singles parametrization is therefore appealing, although it does not make a truncated CC approximation independent of its starting orbitals.
In the ECC formulation,\cite{Arponen1983,ArponenBishopPajanne1987I,ArponenBishopPajanne1987II} the right-hand singles amplitudes $\hT_1$ are accompanied by left-hand amplitudes $\hLam_1$.\cite{PiecuchBartlett1999,FanKowalskiPiecuch2005,FanPiecuch2006}
The latter also contribute to the description of orbital relaxation within ECC.
In the present approximation, the stationary drCCSD ground-state energy is independent of the left-hand amplitudes, but $\hLam_1$ contributes to the transformed Hamiltonian entering the IP/EA-EOM equations and therefore affects the quasiparticle energies.
This makes the simultaneous inclusion of singles on both sides a natural extension of the $GW$-ECC correspondence.

The connection between $GW$ and ECC therefore suggests a different route to reducing the starting-point dependence of $G_0W_0$ by introducing orbital relaxation directly into the CC representation of the ground state.
In the present work, we extend the IP/EA-EOM-drECCD formulation of $G_0W_0$ by including right-hand single-excitation and left-hand single-deexcitation cluster amplitudes, leading to an IP/EA-EOM direct-ring extended coupled-cluster singles and doubles (drECCSD) approximation.
The singles amplitudes are determined in conjunction with the direct-ring doubles amplitudes, allowing orbital relaxation to respond to electron correlation while retaining the ring structure responsible for RPA screening.
The resulting ECC-transformed Hamiltonian is then used within the IP/EA-EOM framework to obtain charged excitation energies.
Unlike renormalized-singles approaches, which modify the reference propagator while preserving the occupied-virtual partition, the present formulation explicitly couples occupied-virtual singles amplitudes to the direct-ring doubles amplitudes within the correlated ground-state equations.
This provides a CC-based route to incorporating orbital relaxation into a $GW$-like quasiparticle treatment without requiring iterative updates of the Green's function. We investigate whether this additional flexibility reduces the sensitivity to HF and KS starting points and assess its impact on the accuracy of core and valence IPs, for which extensive molecular benchmarks and related $GW$ developments are available.\cite{BrunevalDattaniVanSetten2021,marie2024val,GolzeWilhelmVanSettenRinke2018,GolzeKellerRinke2020}

\section{Theory}

\subsection{The $G_0W_0$ approximation}
\label{sec:G0W0}
Hedin's equations form a closed set of coupled equations for the one-body Green's function $G$,
\begin{subequations}
\begin{align}
    \label{eq:G}
    G(12) &= G_\text{H}(12) + G_\text{H}(13) \Sigma_\xc(34) G(42)
    \\
    \label{eq:Gamma}
    \Gamma(123) &= \delta(12) \delta(13)
    + \frac{\delta \Sigma_\xc(12)}{\delta G(45)} G(46) G(75) \Gamma(673)
    \\
    \label{eq:P}
    P(12) &= -\ii G(13) G(41) \Gamma(342)
    \\
    \label{eq:W}
    W(12) &= v(12) + v(13) P(34) W(42)
    \\
    \label{eq:Sigma}
    \Sigma_\xc(12) &= \ii G(14) W(1^+3) \Gamma(423)
\end{align}
\end{subequations}
where $G_\text{H}$ is the Hartree Green's function, $\Gamma$ is the irreducible vertex, $P$ is the irreducible polarizability, $W$ is the screened Coulomb interaction, and $\Sigma_\xc$ is the exchange-correlation self-energy.
The numerical label $1 \equiv \{t_1,\sigma_1,\br_1\} \equiv \{t_1,\bx_1\}$ collects the time, spin, and spatial coordinates. 
Repeated numerical labels imply integration over time and space and summation over spin.
The superscript $^+$ denotes an infinitesimal forward time shift.

The $GW$ approximation is obtained by replacing the vertex by
\begin{equation}
    \Gamma(123) = \delta(12)\delta(13)
\end{equation}
which yields
\begin{subequations}
\begin{align}
    P(12) &= -\ii G(12)G(21)
    \\
    \Sigma_\xc(12) &= \ii G(12)W(1^+2)
\end{align}
\end{subequations}
Solving these equations together with Dyson's equation [see Eq.~\eqref{eq:G}] gives fully self-consistent $GW$.
In the one-shot $G_0W_0$ approximation, instead, the Green's function and screened interaction are constructed from a fixed mean-field reference, usually obtained from Hartree--Fock (HF) or Kohn--Sham (KS) theory.
The corresponding non-interacting, time-ordered Green's function is
\begin{equation}
    \label{eq:G0}
    G_0(\bx_1,\bx_2;\omega)
    = \sum_i \frac{\phi_i^0(\bx_1)\phi_i^{0*}(\bx_2)}{\omega-\epsilon_i^0-\ii0^+}
    + \sum_a \frac{\phi_a^0(\bx_1)\phi_a^{0*}(\bx_2)}{\omega-\epsilon_a^0+\ii0^+}
\end{equation}
where $\phi_p^0$ and $\epsilon_p^0$ are the starting orbitals and orbital energies.
Throughout, $i,j,k,\ldots$ denote occupied orbitals, $a,b,c,\ldots$ virtual orbitals, and $p,q,r,\ldots$ general orbitals.

The screening is determined by the direct random-phase approximation (dRPA) eigenvalue problem
\begin{equation} \label{eq:casida}
    \begin{pmatrix}
        \bA & \bB
        \\
        -\bB^* & -\bA^*
    \end{pmatrix}
    \begin{pmatrix}
        \bX
        \\
        \bY
    \end{pmatrix}_\mu
    = \Omega_\mu
    \begin{pmatrix}
        \bX
        \\
        \bY
    \end{pmatrix}_\mu
\end{equation}
with
\begin{subequations}
\begin{align}
    A_{ai,bj} &= \delta_{ij}\delta_{ab}(\epsilon_a^0-\epsilon_i^0) + \ERI{ja}{bi}
    \\
    B_{ai,bj} &= \ERI{ab}{ij}
\end{align}
\end{subequations}
Here $\ERI{pq}{rs} \equiv \braket{pq}{rs}$ denotes a two-electron integral in Dirac notation, and $X_{i,\mu}^{a}$ and $Y_{a,\mu}^{i}$ are the components of the dRPA eigenvectors.
The index $\mu$ runs over the excitation solutions $\Omega_\mu$, with the eigenvectors normalized according to $\bX^\dagger\bX-\bY^\dagger\bY=\bI$ for a stable dRPA problem. 
The associated screened two-electron integrals are
\begin{subequations} \label{eq:scr_2int}
\begin{align}
    \sERI{pq}{\mu} &= \sum_{ia} \qty[\ERI{pa}{qi} x^a_{i,\mu} + \ERI{pi}{qa} y^i_{a,\mu}]
    \\
    \sERI{\mu}{pq} &= \sum_{ia} \qty[\ERI{qi}{pa} x^{a*}_{i,\mu} + \ERI{qa}{pi} y^{i*}_{a,\mu}]
\end{align}
\end{subequations}

Separating the $GW$ self-energy into its static exchange and dynamical correlation contributions, $\Sigma_\xc^{G_0W_0}=\Sigma_\text{x}^{G_0W_0}+\Sigma_\text{c}^{G_0W_0}$, the latter admits the spectral representation
\begin{equation} \label{eq:SigmaGW}
    \Sigma_{\text{c},pq}^{G_0W_0}(\omega)
    = \sum_{i\mu} \frac{\sERI{pi}{\mu}\sERI{\mu}{qi}}{\omega-\epsilon_i^0+\Omega_\mu-\ii0^+}
    + \sum_{a\mu} \frac{\sERI{ap}{\mu}\sERI{\mu}{aq}}{\omega-\epsilon_a^0-\Omega_\mu+\ii0^+}
\end{equation}
The dRPA problem and the self-energy are evaluated using the same starting orbitals and orbital energies. 
Here, $0^+$ denotes a positive infinitesimal specifying the time-ordering prescription for the self-energy.

Neglecting off-diagonal self-energy elements in the starting orbital basis, the quasiparticle energies are obtained by solving the diagonal Dyson equation
\begin{equation} \label{eq:workingGW}
    \epsilon_p^{G_0W_0}
    = \epsilon_p^0
    + \Sigma_{\text{x},pp}^{G_0W_0}
    - (v_\xc^0)_{pp}
    + \Re[\Sigma_{\text{c},pp}^{G_0W_0}(\epsilon_p^{G_0W_0})]
\end{equation}
where $v_\xc^0$ denotes the exchange-correlation potential already included in the mean-field reference.
For an HF starting point, the exchange terms cancel, and this reduces to
\begin{equation}
    \epsilon_p^{G_0W_0}
    = \epsilon_p^\text{HF}
    + \Re[\Sigma_{\text{c},pp}^{G_0W_0}(\epsilon_p^{G_0W_0})]
\end{equation}
For a KS starting point, the first three terms in Eq.~\eqref{eq:workingGW} reduce to the diagonal elements of the Fock matrix constructed from the reference density.
Since $f_{pp}=\epsilon_p^\text{HF}$ for an HF reference, the quasiparticle equation can be written in a unified form for both HF and KS starting points,
\begin{equation}
    \epsilon_p^{G_0W_0}
    = f_{pp}
    + \Re[\Sigma_{\text{c},pp}^{G_0W_0}
    (\epsilon_p^{G_0W_0})]
\end{equation}
Because $G_0$ and $W_0$ remain fixed throughout the calculation, the resulting quasiparticle energies retain a dependence on the initial mean-field reference.

\subsection{Connection to extended coupled-cluster theory}
\label{sec:ECCGW}

As recently demonstrated, the $G_0W_0$ approximation can be
reformulated within the extended coupled-cluster (ECC)
framework.\cite{ToelleEtAl2026}
In the doubles approximation (ECCD), the doubly
similarity-transformed Hamiltonian reads
\begin{equation}
    \wtH_\text{ECCD}
    =
    e^{\hLam_2} e^{-\hT_2}
    \hH_\text{N}
    e^{\hT_2} e^{-\hLam_2}
\end{equation}
where $\hT_2$ and $\hLam_2$ are the doubles excitation
and deexcitation operators, respectively, and
$\hH_\text{N}$ is the normal-ordered Hamiltonian.

Applying the direct-ring approximation and the associated
quasibosonic representation leads to the drECCD formulation.
Because the resulting bosonic Hamiltonian is quadratic,
the first similarity transformation truncates at second order,
\begin{equation}
    \widetilde{H}_\ldr
    =
    \hH_\text{N}
    + \comm{\hH_\text{N}}{\hT_2}
    + \frac{1}{2}
    \comm{\comm{\hH_\text{N}}{\hT_2}}{\hT_2}
\end{equation}
where the commutators are understood within the direct-ring
quasibosonic approximation.
The second similarity transformation gives
\begin{equation}
\label{eq:H_lamdar}
    \wtH_\ldr
    =
    \widetilde{H}_\ldr
    + \comm{\hLam_2}{\widetilde{H}_\ldr}
    + \frac{1}{2}
    \comm{\hLam_2}{
        \comm{\hLam_2}{\widetilde{H}_\ldr}
    }
\end{equation}
The last term vanishes once the right-amplitude equations
are satisfied, so that the transformed Hamiltonian is
effectively linear in $\hLam_2$ at the stationary solution.

The corresponding drECCD energy functional is
\begin{equation}
\label{eq:drCCD_en}
\begin{split}
    E_\text{drECCD}
    &=
    \mel{\Phi_0}{\wtH_\ldr}{\Phi_0}
    \\
    &=
    \frac{1}{2}\Tr(\bB^*\bT)
    + \frac{1}{2}\Tr(\bLam 
    \boldsymbol{\mathcal{R}})
\end{split}
\end{equation}
where $\Phi_0$ is the reference determinant, $T_{ai,bj}=t_{ij}^{ab}$, and
$\Lambda_{ai,bj}=\lambda_{ab}^{ij}$ are the right- and
left-hand doubles amplitudes, respectively, and
\begin{equation}
\label{eq:drECCD_Riccati}
    \boldsymbol{\mathcal{R}}
    =
    \bB
    + \bA\bT
    + \bT\bA^*
    + \bT\bB^*\bT
\end{equation}
The amplitudes are determined by imposing stationarity
of the energy functional,
\begin{align}
    \pdv{E_\text{drECCD}}{\lambda_{ab}^{ij}} &= 0
    &
    \pdv{E_\text{drECCD}}{t_{ij}^{ab}} &= 0
\end{align}
The first condition yields the Riccati equation
$\boldsymbol{\mathcal{R}}=\bO$, which is equivalent to the dRPA eigenvalue
problem of Eq.~\eqref{eq:casida}.
The second condition determines the left-hand amplitudes
and becomes linear in $\bLam$ once $\bT$ is known.
Consequently, the right- and left-hand amplitude equations
can be solved sequentially.
At stationarity, the second term in
Eq.~\eqref{eq:drCCD_en} vanishes, recovering the dRPA
plasmon formula for the correlation energy.
The correspondence with dRPA can be made explicit through
the relations
\begin{align}
    \bT
    &=
    \bY^* (\bX^*)^{-1}
    &
    \bLam
    &=
    \bY (\bX-\bT\bY)^{-1}
\end{align}
where $\bX$ and $\bY$ are the dRPA excitation and
deexcitation amplitudes introduced previously.

The connection with $G_0W_0$ extends beyond the
ground-state correlation energy.
Indeed, applying the equation-of-motion (EOM) formalism
to the drECCD Hamiltonian allows one to recover the
$G_0W_0$ quasiparticle energies.
Within the quasibosonic formulation, the EOM operator
spans the one-hole (1h), two-hole--one-particle (2h1p),
one-particle (1p), and two-particle--one-hole (2p1h) 
configurations.
In this combined quasibosonic EOM representation, the 1p and 2p1h contributions are deexcitation components, rather than electron-addition operators acting on the reference determinant. They must be retained alongside the 1h and 2h1p terms to preserve the coupling of the two branches in this representation. 
The complete IP/EA-EOM operator thus reads
\begin{equation}
\begin{split}
    \hR
    &=
    \sum_i r_i \ani{i}
    + \sum_{ijb} r_{ji}^b
      \cre{b}\ani{j}\ani{i}
    \\
    &+
    \sum_a r_a \ani{a}
    + \sum_{ajb} r_{ba}^j
      \cre{j}\ani{b}\ani{a}
\end{split}
\end{equation}
Here, the composite operator strings represent the
corresponding quasibosonic configurations rather than
fully antisymmetrized fermionic excitation operators.
The resulting EOM eigenvalue problem reads
\begin{equation}
\label{eq:EOM_right}
    \mel{\Phi_0}{
        \comm{\hat{c}_I^\dagger}{
            \comm{\wtH_\ldr}{\hR}
        }_+
    }{\Phi_0}
    =
    \epsilon^\text{EOM}
    \mel{\Phi_0}{
        \comm{\hat{c}_I^\dagger}{\hR}_+
    }{\Phi_0}
\end{equation}
where $\comm{\cdot}{\cdot}_+$ denotes the
anticommutator and $\hat{c}_I$ represents an operator
belonging to the EOM configuration space.

Within the diagonal approximation, the one-particle
component associated with orbital $p$ is restricted to
\begin{equation}
\label{eq:diagonal_r}
    \hR_{1,p}
    \approx
    \sum_q r_q \ani{q}\delta_{pq}
    =
    r_p\ani{p}
\end{equation}
The corresponding EOM eigenvalues then reproduce the
$G_0W_0$ quasiparticle energies according to
\begin{equation}
    \epsilon_p^{\text{EOM}}
    =
    -\epsilon_p^{G_0W_0}
\end{equation}
This equivalence provides the foundation for extending
the $GW$ approximation using established CC techniques,
as discussed in the following section.

\subsection{Orbital relaxation within drECCD}
\label{sec:orbital_relaxation}

Building on the connection between $G_0W_0$ and ECC established
in the previous section, we seek to alleviate the starting-point
dependence of $G_0W_0$ without introducing a self-consistency cycle
for the Green's function or increasing the formal scaling of the
ground-state equations beyond $\order*{K^6}$, where $K$ denotes
the number of orbitals.
To this end, we introduce single excitations into the right-hand cluster operator and single deexcitations into its left-hand counterpart, allowing orbital relaxation to be determined alongside the direct-ring doubles amplitudes.
This strategy is motivated by the role of singles in conventional
CC approximations, including CC2, where they account for orbital
relaxation effects absent from second-order Møller--Plesset
perturbation theory (MP2). \cite{Christiansen1995}

We start from the ECC singles and doubles (ECCSD) Hamiltonian
\begin{equation}
    \wtH_\text{ECCSD}
    =
    e^{\hLam_1+\hLam_2}
    e^{-\hT_1-\hT_2}
    \hH_\text{N}
    e^{\hT_1+\hT_2}
    e^{-\hLam_1-\hLam_2}
\end{equation}
Truncating the Baker--Campbell--Hausdorff expansion of the
left-hand similarity transformation at first order in the
deexcitation operators gives
\begin{equation}
    \wtH_{\lambda\text{CCSD}}
    =
    \widetilde{H}
    +
    \comm{\hLam_1+\hLam_2}{\widetilde{H}}
\end{equation}
where
\begin{equation}
    \widetilde{H}
    =
    e^{-\hT_1-\hT_2}
    \hH_\text{N}
    e^{\hT_1+\hT_2}
\end{equation}
This approximation recovers the conventional biorthogonal
formulation of CCSD and retains its formal $\order*{K^6}$
scaling.
In particular, the right-hand amplitude equations are
independent of the left-hand amplitudes and read
\begin{align}
    \mel{\Phi_0}{
        \pdv{\wtH_{\lambda\text{CCSD}}}{\lambda_J}
    }{\Phi_0}
    &=
    0
    &
    \mel{\Phi_J}{\widetilde{H}}{\Phi_0}
    &=
    0
\label{eq:t-lCCSD}
\end{align}
where $J$ spans the single- and double-excitation spaces.
The corresponding left-hand amplitudes are determined from
\begin{subequations}
\begin{align}
    \mel{\Phi_0}{
        \pdv{\wtH_{\lambda\text{CCSD}}}{t_J}
    }{\Phi_0}
    &=
    0
    \\
    \mel{\Phi_0}{
        (1+\hLam_1+\hLam_2)
        e^{-\hT_1-\hT_2}
        \comm{\hH_\text{N}}{\hat c_J}
        e^{\hT_1+\hT_2}
    }{\Phi_0}
    &=
    0
\end{align}
\end{subequations}
with $\hat c_J$ the corresponding excitation operator.

To recover the direct-ring structure underlying $G_0W_0$,
we next apply the quasiboson approximation to the doubles
equations. \cite{SchuckBook,ScuseriaHendersonSorensen2008}
In this approximation, particle-hole excitations are treated
as independent bosonic degrees of freedom, and contributions
that cannot be expressed in terms of these excitations are
discarded.
Schematically, the double-excitation operators are replaced by
\begin{equation}
    \cre{a}\cre{b}\ani{j}\ani{i}
    \longrightarrow
    b_{ai}^\dagger b_{bj}^\dagger
\end{equation}
where $b_{ai}^\dagger$ creates a quasibosonic particle-hole
excitation from orbital $i$ to orbital $a$.
The additional direct approximation neglects exchange
contributions in the corresponding particle-hole interaction.

It is convenient to separate the singles transformation by
introducing the partially similarity-transformed Hamiltonian
\begin{equation}
    \widetilde{H}_1
    =
    e^{-\hT_1}\hH_\text{N}e^{\hT_1}
\end{equation}
We denote by $\wF_{pq}$ and $\wW_{pqrs}$ the one- and
two-body intermediates of $\widetilde{H}_1$, respectively.
More precisely, they are defined for general orbital indices as the coefficients of the normal-ordered operators in
\begin{equation}
  \widetilde{H}_1 = E_1 + \sum_{pq}\wF_{pq}\NO{\cre{p}\ani{q}} + \frac14\sum_{pqrs}\wW_{pqrs}\NO{\cre{p}\cre{q}\ani{s}\ani{r}}
\end{equation}
where $\wW_{pqrs}$ is antisymmetric. 
The occupied-virtual matrix elements entering the singles equations are
\begin{align}
    \wF_{ai}
    & = \mel{\Phi_i^a}{\widetilde{H}_1}{\Phi_0}
    &
    \wF_{ia}
    & = \mel{\Phi_0}{\widetilde{H}_1}{\Phi_i^a}
\end{align}

The right-hand CCSD equations can then be rewritten as
\begin{subequations}
\begin{align}
    \mel{\Phi_i^a}{
        e^{-\hT_2}\widetilde{H}_1e^{\hT_2}
    }{\Phi_0}
    &=
    0
    \label{eq:t1-lCCSD}
    \\
    \mel{\Phi_{ij}^{ab}}{
        e^{-\hT_2}\widetilde{H}_1e^{\hT_2}
    }{\Phi_0}
    &=
    0
    \label{eq:t2-lCCSD}
\end{align}
\end{subequations}
Applying the quasiboson approximation to
Eq.~\eqref{eq:t2-lCCSD} yields a Riccati equation for
the ring-doubles amplitudes,
\begin{equation}
\label{eq:rCCSD_riccati}
    \widetilde{B}_{ai,bj}
    +
    \sum_{ld}
    P_{ai,bj}^+
    \widetilde{A}_{ai,dl}t_{lj}^{db}
    +
    \sum_{lmde}
    t_{il}^{ad}
    B^*_{dl,em}
    t_{mj}^{eb}
    =
    0
\end{equation}
or, equivalently,
\begin{equation}
    \widetilde{\bB}
    +
    \widetilde{\bA}\bT
    +
    \bT\widetilde{\bA}^{T}
    +
    \bT\bB^*\bT
    =
    \bO
\end{equation}
Here, the symmetrization operator exchanges the particle-hole
pairs $(ai)$ and $(bj)$,
\begin{equation}
    P_{ai,bj}^+
    \widetilde{A}_{ai,dl}t_{lj}^{db}
    =
    \widetilde{A}_{ai,dl}t_{lj}^{db}
    \\
    +
    \widetilde{A}_{bj,dl}t_{li}^{da}
\end{equation}
and the partially transformed matrices are defined as
\begin{subequations}
\begin{align}
    \widetilde{B}_{ai,bj}
    &=
    \mel{\Phi_{ij}^{ab}}{\widetilde{H}_1}{\Phi_0}
    \\
    B^*_{ai,bj}
    &=
    \mel{\Phi_0}{\widetilde{H}_1}{\Phi_{ij}^{ab}}
    =
    \mel{\Phi_0}{\hH_\text{N}}{\Phi_{ij}^{ab}}
    \\
    \widetilde{A}_{ai,bj}
    &=
    \mel{\Phi_i^a}{
        \widetilde{H}_1-E_1
    }{\Phi_j^b}
\end{align}
\end{subequations}
With this notation, the forward pair-creation block $\widetilde{\bB}$ depends on $T_1$, whereas the pair-annihilation block remains equal to the untransformed $\bB^*$. This distinction is also used in the generalized RPA matrix below.
The subtraction of the reference expectation value
\begin{equation}
    E_1 = \mel{\Phi_0}{\widetilde{H}_1}{\Phi_0}
\end{equation}
ensures that $\widetilde{\bA}$ contains only the connected
contributions to the particle-hole excitation matrix.
Equivalently, its elements can be expressed as
\begin{equation}
    \widetilde{A}_{ai,bj}
    =
    \mel{\Phi_i^a}{
        \comm{\widetilde{H}_1}{\cre{b}\ani{j}}
    }{\Phi_0}
\end{equation}

A crucial distinction arises when considering the singles
equation [Eq.~\eqref{eq:t1-lCCSD}].
The quasiboson approximation introduced above for the
doubles equation [Eq.~\eqref{eq:t2-lCCSD}] could also be
applied to the singles equation. However, doing so would
discard all but one of the terms coupling the doubles
amplitudes to the singles residual. The only remaining
contribution would be
$\sum_{ck}t_{ik}^{ac}\wF_{kc}$,
involving the occupied-virtual block of the partially
transformed one-body operator.
For an HF reference, this approximation admits the
solution $t_i^a=0$, thereby suppressing the orbital
relaxation effects that we seek to describe.
We therefore treat the two equations differently.
In particular, the fermionic Wick contractions of the singles
residual [Eq.~\eqref{eq:t1-lCCSD}] are retained,
whereas the doubles residual
[Eq.~\eqref{eq:t2-lCCSD}] is evaluated in the
quasibosonic direct-ring approximation.
The resulting singles residual has the same structure
as in conventional CCSD,
\begin{equation}
\label{eq:singlesCC}
    0=
    \wF_{ai}
    +
    \sum_{ck}\tilde{t}_{ik}^{ac}\wF_{kc}
    \\
    +
    \frac12 \sum_{cdk}\wW_{akcd}t_{ik}^{cd}
    -
    \frac12 \sum_{ckl}\wW_{klic}t_{kl}^{ac}
\end{equation}
The last two terms retain the coupling between the singles
and doubles amplitudes through the transformed two-electron
intermediates.
This coupling is essential for allowing the singles amplitudes
to respond to electron correlation.

An important consequence of the ring approximation is that
the doubles amplitudes no longer possess the usual fermionic
antisymmetry.
Instead, they satisfy the pair symmetry
$t_{ij}^{ab}=t_{ji}^{ba}$,
but generally
$t_{ij}^{ab}\neq-t_{ij}^{ba}$.
To correctly account for the contributions of these amplitudes in the fermionic singles equations,
we introduce the explicitly antisymmetrized combination
\begin{equation}
\label{eq:anti_t2}
    \tilde{t}_{ij}^{ab}
    =
    \frac12
    \left(
        t_{ij}^{ab}-t_{ij}^{ba}
    \right)
\end{equation}
This antisymmetrized intermediate is used to evaluate
the contribution of the one-body part of the partially
similarity-transformed Hamiltonian to the fermionic singles
residual according to Wick's theorem.
It does not impose fermionic antisymmetry on the
underlying ring-doubles amplitudes.

The direct approximation introduces an additional subtlety
because the partially transformed two-electron intermediates
$\wW_{pqrs}$ contain contributions involving both Coulomb
and exchange integrals.
We construct their direct counterparts $\wWd_{pqrs}$ using
the non-antisymmetrized integrals $\ERI{pq}{rs}$ rather than
$\aERI{pq}{rs}=\ERI{pq}{rs}-\ERI{pq}{sr}$, while imposing
the reconstruction condition
\begin{equation}
    \wW_{pqrs}
    =
    \wWd_{pqrs}-\wWd_{pqsr}
    =
    \wWd_{pqrs}-\wWd_{qprs}
\end{equation}
In particular, contributions involving the
$\ERI{ia}{bj}$ and $\ERI{ai}{bj}$ blocks require care
because both index orderings appear in the singles
transformation.
For the same-spin blocks, we employ an equal-weight
average of the corresponding contributions, introducing
a factor of $1/2$.
This prescription affects the
$\wWd_{abij}$, $\wWd_{abci}$, and $\wWd_{kaij}$
intermediates.
Their explicit expressions are given in
Appendix~\ref{app:Wd_pqrs}. 
The factors of $1/2$ associated with this equal-weight prescription are identified explicitly in Appendix~\ref{app:Wd_pqrs}. 

The spin structure of the antisymmetrized ring amplitudes
must also be considered.
For the $\alpha\alpha\alpha\alpha$ and
$\beta\beta\beta\beta$ blocks, we use the definition
in Eq.~\eqref{eq:anti_t2}.
For the mixed-spin block, however, the direct
approximation excludes the corresponding spin-flip
amplitudes.
Consistently with the spin summation employed in the
direct-ring correlation energy, we adopt
$\tilde{t}_{\ i_\alpha j_\beta}^{\ a_\alpha b_\beta}
= t_{i_\alpha j_\beta}^{a_\alpha b_\beta}$. 
This prescription retains the mixed-spin contribution
without introducing the additional factor of $1/2$
associated with Eq.~\eqref{eq:anti_t2}. 

Combining these approximations gives the following
working equations for the right-hand amplitudes:
\begin{subequations}
\begin{align}
    0&=
    \wF_{ai}
    +
    \sum_{ck}\tilde{t}_{ik}^{ac}\wF_{kc}
    +
    \sum_{cdk}\wWd_{akcd}\tilde{t}_{ik}^{cd}
    -
    \sum_{ckl}\wWd_{klic}\tilde{t}_{kl}^{ac}
    \label{eq:singlesrCCSD}
    \\
    0&=
    \wWd_{abij}
    +
    \sum_{ld}
    P_{ai,bj}^+
    \left(
        \delta_{il}\wF_{ad}
        -
        \delta_{ad}\wF_{li}
        +
        \wWd_{ladi}
    \right)t_{lj}^{db}
    +
    \sum_{lmde}
    t_{im}^{ae}\ERI{ml}{ed}t_{lj}^{db}
    \label{eq:doublesrCCSD}
\end{align}
\end{subequations}

In Eq.~\eqref{eq:singlesrCCSD}, the use of the
antisymmetrized contraction amplitudes
$\tilde{t}_{ij}^{ab}$ allows the antisymmetrized
two-electron integrals to be replaced by their direct
counterparts, thereby absorbing the explicit factors
of $1/2$ appearing in Eq.~\eqref{eq:singlesCC}.

The corresponding left-hand equations are obtained by
requiring stationarity of the approximate CC Lagrangian
with respect to the singles and doubles amplitudes.
The multipliers $\lambda_a^i$ and $\lambda_{ab}^{ij}$
are associated with the residuals in
Eqs.~\eqref{eq:singlesrCCSD} and
\eqref{eq:doublesrCCSD}, respectively.
The resulting equations read
\begin{equation}
\begin{split}
    0=
    \wF_{ia}
    &+
    \sum_{dl}
    \left(
        \wWd_{idal}
        +
        \sum_{ck}
        \tilde{t}_{lk}^{dc}\ERI{ik}{ac}
    \right)\lambda_d^l
    +
    \sum_d
    \left(
        \wF_{da}
        -
        \sum_{ckl}
        \tilde{t}_{lk}^{dc}\ERI{lk}{ac}
    \right)\lambda_d^i
    \\
    &-
    \sum_l
    \left(
        \wF_{il}
        +
        \sum_{cdk}
        \tilde{t}_{kl}^{cd}\ERI{ik}{dc}
    \right)\lambda_a^l
    -
    \sum_{ckl}
    \wWd_{iclk}\lambda_{ac}^{lk}
    +
    \sum_{cdk}
    \wWd_{dcak}\lambda_{dc}^{ik}
    \\
    &+
    \sum_{cdekm}
    \wW_{cida}t_{km}^{ce}\lambda_{de}^{km}
    +
    \sum_{ceklm}
    \wW_{kila}t_{km}^{ce}\lambda_{ce}^{lm}
    -
    \sum_{cekm}
    \wF_{ic}t_{km}^{ce}\lambda_{ae}^{km}
    \\
    &+
    \sum_{cekm}
    \wF_{ka}t_{km}^{ce}\lambda_{ce}^{im}
    -
    \sum_{ceklm}
    \wWd_{iklc}t_{km}^{ce}\lambda_{ae}^{lm}
    +
    \sum_{cdekm}
    \wWd_{ckad}t_{km}^{de}\lambda_{ce}^{im}
\end{split}
\end{equation}
and
\begin{equation}
\begin{split}
    0=
    \ERI{ij}{ab}
    &+
    P_{ai,jb}^+
    \left(
        -\sum_k\wF_{jk}\lambda_{ab}^{ik}
        +
        \sum_c\wF_{cb}\lambda_{ac}^{ij}
    \right)
    \\
    &+
    P_{ai,jb}^+
    \sum_{ck}
    \left(
        \wWd_{jcbk}
        +
        \sum_{dl}
        \ERI{jl}{bd}t_{lk}^{dc}
    \right)\lambda_{ac}^{ik}
    \\
    &+
    P_{ab}^-P_{ij}^-
    \wF_{jb}\lambda_a^i
    \\
    &+
    P_{ai,jb}^+
    \left(
        -\sum_k\wWd_{ijkb}\lambda_a^k
        +
        \sum_c\wWd_{cjab}\lambda_c^i
    \right)
\end{split}
\end{equation}
Here, $P_{pq}^-$ denotes antisymmetrization with respect
to the corresponding orbital indices. 
For example,
\begin{equation}
    P_{ab}^-P_{ij}^-
    \wF_{jb}\lambda_a^i
    =
    \wF_{jb}\lambda_a^i
    -
    \wF_{ib}\lambda_a^j
    -
    \wF_{ja}\lambda_b^i
    +
    \wF_{ia}\lambda_b^j
\end{equation}

We refer to the resulting approximation as direct-ring
coupled-cluster singles and doubles (drCCSD).
Its ground-state correlation energy is
\begin{equation}
\label{eq:drCCSD_energy}
    E_\text{drCCSD}
    =
    \sum_{ai}f_{ia}t_i^a
    +
    \frac12
    \sum_{abij}
    \left[
        P_{ij}^-(t_i^at_j^b)
        +
        t_{ij}^{ab}
    \right]\ERI{ij}{ab}
\end{equation}

\subsection{Treating the drCCSD approximation using an RPA formulation}
Because the drCCSD doubles-amplitude equations
[Eqs.~\eqref{eq:rCCSD_riccati} and \eqref{eq:doublesrCCSD}]
retain the Riccati structure of the drCCD equations,
they can be recast as a generalized RPA eigenvalue problem.
Following the procedure of Ref.~\onlinecite{ScuseriaHendersonSorensen2008},
we introduce the Casida-like eigenvalue equation
\begin{equation}
    \begin{pmatrix}
        \widetilde{\bA}^{\dagger} & \bB \\
        -\widetilde{\bB}^{*} & -\widetilde{\bA}^{*}
    \end{pmatrix}
    \begin{pmatrix}
        \widetilde{\bX} \\
        \widetilde{\bY}
    \end{pmatrix}
    =
    \begin{pmatrix}
        \widetilde{\bX} \\
        \widetilde{\bY}
    \end{pmatrix}
    \widetilde{\bOm}
\end{equation}
from which the doubles-amplitude matrix
$({\bT_2})_{ai,bj}=t_{ij}^{ab}$ can be reconstructed as
\begin{equation}
    \bT_2
    =
    \widetilde{\bY}^{*}
    \qty(\widetilde{\bX}^{*})^{-1}
\end{equation}
This relation reproduces the drCCSD Riccati equation,
provided that $\widetilde{\bX}$ is invertible.

Unlike conventional RPA, the present singles-RPA formulation
involves two distinct coupling matrices,
$\bB$ and $\widetilde{\bB}$, connecting the
excitation and deexcitation sectors.
Nevertheless, because both matrices are symmetric,
the generalized RPA matrix retains the
excitation/deexcitation eigenvalue pairing:
if $\widetilde{\Omega}_\mu$ is an eigenvalue,
so is $-\widetilde{\Omega}_\mu$.
A proof of this property is given in
Appendix~\ref{app:sym_sRPA}.
However, the corresponding eigenvectors no longer exhibit,
in general, the usual conjugation symmetry of conventional RPA.
In particular, the deexcitation eigenvectors cannot
necessarily be constructed from their excitation
counterparts by exchanging and complex conjugating
their excitation and deexcitation components.

Finally, although the doubles amplitudes can be reconstructed
from the singles-RPA eigenvectors, the singles amplitudes
must still be determined from their own equation,
Eq.~\eqref{eq:singlesrCCSD}, coupled to the doubles equations.
Since the drCCSD correlation energy depends explicitly on
both sets of amplitudes [Eq.~\eqref{eq:drCCSD_energy}],
the singles-RPA eigenvalue problem alone does not determine
this energy.
Furthermore, no analogous relationship has been established
for the left-hand $\lambda$ amplitudes.
In particular, their connection to the block diagonalization
of the generalized RPA matrix remains unclear, unlike in the
RPA/drECCD correspondence.\cite{KitsarasToelleLoos2026,ToelleEtAl2026}
We therefore do not establish a plasmon formula expressing
the drCCSD correlation energy solely in terms of the
singles-RPA excitation energies.

\subsection{Orbital-relaxed $G_0W_0$ quasiparticle energies}
\label{sec:sG0W0}

Although the drCCSD doubles-amplitude equations can be
recast as a generalized RPA eigenvalue problem, the latter
is not equivalent to a conventional RPA calculation of the
polarizability in an orbital basis relaxed by the $t_i^a$
amplitudes. Consequently, a direct derivation of the
corresponding quasiparticle energies from Hedin's equations
is not straightforward.

Instead, we follow the EOM formulation introduced in
Sec.~\ref{sec:ECCGW} and obtain the orbital-relaxed
quasiparticle energies from
\begin{equation}
\label{eq:sEOM_right}
  \mel{\Phi_0}{
    \comm{\hat{c}_I^\dagger}{
      \comm{\wtH_{\lambda\text{-drCCSD}}}{\hR}
    }_+
  }{\Phi_0}
  =
  \epsilon^\text{EOM}
  \mel{\Phi_0}{
    \comm{\hat{c}_I^\dagger}{\hR}_+
  }{\Phi_0}
\end{equation}
The resulting working equations closely resemble those
of Eq.~\eqref{eq:EOM_right}, allowing us to retain the
same computational framework. In particular, within the
diagonal approximation of Eq.~\eqref{eq:diagonal_r},
the quasiparticle problem can be solved using a linear
solver rather than a Davidson eigensolver.

In practice, the inclusion of singles requires only
a few modifications to the existing EOM implementation:
\begin{itemize}
  \item
  Contributions involving the Fock matrix $f_{pq}$
  and two-electron integrals $\ERI{pq}{rs}$ contracted
  with $\hR$, or with both $\hT_2$ and $\hR$,
  are replaced by the corresponding intermediates
  $\wFF_{pq}$ and $\wVd_{pqrs}$, respectively.
  These arise from the
  $(1+\hLam_1)\widetilde{H}_1$ part of the
  transformed Hamiltonian and are defined in
  Appendix~\ref{app:Vd_pqrs}.

  \item
  Contributions involving two-electron integrals
  $\ERI{pq}{rs}$ contracted with both $\hLam_2$
  and $\hR$ are evaluated using the transformed
  direct integrals $\wWd_{pqrs}$
  arising from $\widetilde{H}_1$.
\end{itemize}
We refer to the resulting quasiparticle energies as
singles-$G_0W_0$ (s$G_0W_0$) quasiparticle energies.

\section{Computational details}

The methods developed in this work have been implemented in
the \textsc{qcumbre} program package.\cite{qcumbre}
The implementation of ring and direct-ring CCSD
[(d)rCCSD] builds upon the existing CC2 and (d)rCCD
implementations.
The right-hand amplitude equations were validated by
recovering the CC2 energies and amplitudes when the
additional (d)rCCD contributions are disabled, and the
(d)rCCD results when the singles amplitudes are set to zero.
The left-hand cluster amplitude equations were independently
validated by verifying that the numerical derivatives of
the drCCSD Lagrangian with respect to the right-hand cluster
amplitudes vanish at convergence.
The s$G_0W_0$ implementation builds upon the existing
IP/EA-EOM formulation of $G_0W_0$, with the modifications
described in Sec.~\ref{sec:sG0W0}.

Unless otherwise stated, the HF reference orbitals were obtained with \textsc{cfour}.~\cite{Matthews_2020,cfour}
For core IPs calculated from total-energy differences, the open-shell core-ionized states were described using unrestricted HF (UHF) references.
We also used \textsc{quack}~\cite{QuAcK} when \textsc{cfour} failed to converge to the target core-ionized state or when the UHF solutions obtained with \textsc{cfour} differed from those reported in Ref.~\onlinecite{marie2026core}.
When multiple HF solutions were obtained for the same target state, we selected the lowest-energy solution.
Cases in which \textsc{quack}~\cite{QuAcK} was used are explicitly identified below.

The PBE0 and PBEh($\alpha$) reference orbitals were obtained with PySCF.~\cite{pyscf1,pyscf2,pyscf3,pyscf4} PBEh($\alpha$) denotes the hybrid PBE functional with a fraction $\alpha$ of exact exchange. 
For the $G_0W_0$@PBEh($\alpha$) core IPs, we used 45\% exact exchange for second-row core-ionized atoms, as proposed in Ref.~\onlinecite{GolzeKellerRinke2020}, and 70\% for third-row core-ionized atoms, as in Ref.~\onlinecite{marie2026core}.

All correlated calculations based on an HF reference, including drCCD, drCCSD, $G_0W_0$, and s$G_0W_0$, were performed with \textsc{qcumbre}.
The s$G_0W_0$ calculations using PBE0 and PBEh($\alpha$) starting points were likewise performed with \textsc{qcumbre}, using the corresponding KS orbitals generated with PySCF.
The $G_0W_0$@PBE0 valence IPs were computed with \textsc{molgw}.~\cite{molgw}

\section{Results}
We consider two molecular benchmark sets, originally assembled for inner- and outer-valence IPs\cite{marie2024val} and core IPs,\cite{marie2026core} respectively. In both cases, we employ the molecular geometries of the original studies.

\begin{squeezetable}
 \begin{table}
   \caption{Valence IPs (in \si{\eV}) for the benchmark set of Ref.~\onlinecite{marie2024val}, calculated in the aug-cc-pVTZ basis set using $G_0W_0$, s$G_0W_0$ with either HF or PBE0 as the starting point, qs$GW$ starting from HF orbitals, as well as FCI. The HOMO$-n$ label identifies the ionized orbital, with $n=0$ denoting the HOMO. Error statistics are evaluated relative to FCI and reported in \si{\eV}: mean signed error (MSE), mean absolute error (MAE), standard deviation of the errors (SDE), and minimum and maximum signed errors (Min and Max).}
   \label{tab:val_ip}
 	\begin{ruledtabular}
     \begin{tabular}{
         lc*{6}{S[table-format=2.3]}
       }
       & \multicolumn{1}{c}{\multirow{2}{*}{HOMO$-n$}} 
       & \multicolumn{1}{c}{$G_0W_0$\rlap{\fnm[1]}}
       & \multicolumn{1}{c}{$G_0W_0$\rlap{\fnm[2]}} 
       & \multicolumn{1}{c}{s$G_0W_0$\rlap{\fnm[2]}} 
       & \multicolumn{1}{c}{s$G_0W_0$\rlap{\fnm[2]}}    
       & \multicolumn{1}{c}{qs$GW$\rlap{\fnm[1]}} 
       & \multicolumn{1}{c}{\multirow{2}{*}{FCI\rlap{\fnm[1]}}}\\ 
       & 
       & \multicolumn{1}{c}{@HF} 
       & \multicolumn{1}{c}{@PBE0} 
       & \multicolumn{1}{c}{@HF} 
       & \multicolumn{1}{c}{@PBE0} 
       & \multicolumn{1}{c}{@HF} 
       & \\\hline 
       \multirow{2}{*}{\ce{BH3}} & 0        & 13.678 & 13.156 & 13.584 & 13.550 & 13.584 & 13.307   \\
 						                     & 1        & 18.685 & 17.684 & 18.584 & 18.544 & 18.544 & 18.310   \\
 			\multirow{2}{*}{\ce{CH4}}  & 0        & 14.753 & 14.155 & 14.625 & 14.615 & 14.621 & 14.377   \\
 			                           & 1        & 23.875 & 22.339 & 23.716 & 23.695 & 23.550 & 23.146   \\
 			\multirow{3}{*}{\ce{H2O}}  & 0        & 12.884 & 12.160 & 12.955 & 12.797 & 12.879 & 12.679   \\
 						                     & 1        & 15.106 & 14.404 & 15.185 & 15.021 & 15.107 & 14.915   \\
 			                           & 2        & 19.129 & 18.556 & 19.196 & 19.106 & 19.188 & 19.024   \\
 			\multirow{2}{*}{\ce{HF}}   & 0        & 16.237 & 15.498 & 16.451 & 16.225 & 16.349 & 16.149   \\
 			                           & 1        & 20.074 & 19.478 & 20.269 & 20.071 & 20.203 & 20.054   \\
 			\multirow{2}{*}{\ce{Ne}}   & 0        & 21.432 & 20.698 & 21.699 & 21.458 & 21.592 & 21.461   \\
 			                           & 1        & 47.950 & 45.967 & 48.227 & 47.975 & 47.560 & 48.306   \\
 			\multirow{3}{*}{\ce{NH3}}  & 0        & 11.201 & 10.512 & 11.142 & 11.046 & 11.094 & 10.901   \\
 			                           & 1        & 16.867 & 16.212 & 16.810 & 16.773 & 16.805 & 16.603   \\
 			                           & 2        & 28.427 & 26.993 & 28.332 & 28.299 & 27.980 & 26.779   \\
 			\multirow{2}{*}{\ce{BeO}}  & 0        & 9.727  & 9.425  & 10.149 & 9.816  & 10.168 & 9.972    \\
 			                           & 1        & 10.996 & 10.248 & 11.402 & 10.950 & 11.241 & 11.056   \\
 			\multirow{2}{*}{\ce{BN}}   & 0        & 11.752 & 11.435 & 11.747 & 11.793 & 11.898 & 11.980   \\
 			                           & 1        & 13.447 & 13.182 & 13.506 & 13.538 & 13.621 & 13.729   \\
 			\ce{C2}                    & 0        & 12.928 & 12.287 & 12.492 & 12.609 & 12.561 & 12.463   \\
 			\multirow{2}{*}{\ce{LiF}}  & 0        & 11.384 & 10.641 & 11.774 & 11.315 & 11.575 & 11.468   \\
 			                           & 1        & 11.915 & 11.181 & 12.306 & 11.896 & 12.119 & 11.933   \\
 			\multirow{3}{*}{\ce{BF}}   & 0        & 11.325 & 10.817 & 11.185 & 11.167 & 11.167 & 11.149   \\
 			                           & 1        & 18.743 & 17.771 & 18.979 & 18.642 & 18.784 & 18.612   \\
 			                           & 2        & 21.567 & 20.572 & 21.790 & 21.505 & 21.585 & 20.906   \\
 			\multirow{3}{*}{\ce{CO}}   & 0        & 14.777 & 13.763 & 14.454 & 14.452 & 14.318 & 13.925   \\
 			                           & 1        & 17.083 & 16.587 & 17.137 & 17.053 & 17.124 & 16.966   \\
 			                           & 2        & 20.300 & 19.238 & 20.405 & 20.235 & 20.191 & 19.638   \\
 			\multirow{3}{*}{\ce{N2}}   & 0        & 16.350 & 15.263 & 16.064 & 16.050 & 15.914 & 15.486   \\
 			                           & 1        & 17.093 & 16.526 & 16.879 & 16.940 & 16.903 & 16.933   \\
 			                           & 2        & 19.862 & 18.496 & 19.627 & 19.609 & 19.425 & 18.692   \\
 			\multirow{4}{*}{\ce{CH2O}} & 0        & 11.380 & 10.645 & 11.398 & 11.180 & 11.241 & 10.899   \\
 			                           & 1        & 14.587 & 14.088 & 14.650 & 14.570 & 14.693 & 14.578   \\
 			                           & 2        & 16.602 & 15.664 & 16.645 & 16.454 & 16.459 & 16.106   \\
 			                           & 3        & 18.015 & 16.840 & 17.912 & 17.913 & 17.821 & 17.107   \\
 			\multirow{2}{*}{\ce{Ar}}   & 0        & 15.711 & 15.158 & 15.686 & 15.686 & 15.633 & 15.613   \\
 			                           & 1        & 31.089 & 28.523 & 31.063 & 31.077 & 30.681 & 29.188   \\
 			\multirow{3}{*}{\ce{F2}}   & 0        & 16.334 & 15.336 & 16.434 & 16.278 & 16.229 & 15.729   \\
 		 	                           & 1        & 19.902 & 18.608 & 20.010 & 19.853 & 19.752 & 18.874   \\
 			                           & 2        & 20.836 & 20.369 & 20.997 & 20.900 & 21.000 & 21.100   \\
 			\multirow{3}{*}{\ce{H2S}}  & 0        & 10.500 & 10.045 & 10.441 & 10.464 & 10.445 & 10.393   \\
 			                           & 1        & 13.614 & 13.077 & 13.557 & 13.557 & 13.530 & 13.411   \\
 			                           & 2        & 15.928 & 15.269 & 15.832 & 15.829 & 15.785 & 15.627   \\
 			\multirow{2}{*}{\ce{HCl}}  & 0        & 12.778 & 12.269 & 12.747 & 12.754 & 12.723 & 12.676   \\
 			                           & 1        & 16.872 & 16.292 & 16.811 & 16.810 & 16.752 & 16.657   \\
 			\multirow{3}{*}{\ce{PH3}}  & 0        & 10.787 & 10.354 & 10.714 & 10.722 & 10.725 & 10.596   \\
 			                           & 1        & 14.099 & 13.491 & 13.997 & 13.960 & 13.954 & 13.745   \\
 			                           & 2        & 21.567 & 19.192 & 21.468 & 21.401 & 21.261 & 19.514   \\
 			\multirow{2}{*}{\ce{SiH4}} & 0        & 13.214 & 12.591 & 13.142 & 13.058 & 13.107 & 12.793   \\
 			                           & 1        & 19.007 & 17.548 & 18.944 & 18.835 & 18.828 & 18.240   \\
 			\multirow{2}{*}{\ce{LiCl}} & 0        & 9.984  & 9.447  & 9.997  & 9.939  & 9.947  & 9.897    \\
 			                           & 1        & 10.690 & 10.132 & 10.700 & 10.653 & 10.642 & 10.577   \\
 			\multirow{4}{*}{\ce{CO2}}  & 0        & 14.221 & 13.479 & 14.169 & 14.032 & 14.054 & 13.733   \\
 			                           & 1        & 18.519 & 17.295 & 18.415 & 18.297 & 18.200 & 17.513   \\
 			                           & 2        & 18.791 & 17.569 & 18.728 & 18.546 & 18.442 & 17.996   \\
 			                           & 3        & 20.150 & 18.781 & 20.114 & 19.918 & 19.776 & 19.254   \\
 			\multirow{3}{*}{\ce{CS}}   & 0        & 12.378 & 11.249 & 11.949 & 11.935 & 11.775 & 11.300   \\
 			                           & 1        & 12.907 & 12.551 & 12.825 & 12.878 & 12.852 & 12.882   \\
 			                           & 2        & 17.976 & 16.554 & 17.838 & 17.865 & 17.688 & 17.920   \\
 						\hline\hline                                                    
      MSE                        &          & 0.399  & -0.469 & 0.397  &  0.298 & 0.290  &           \\
 			MAE                        &          & 0.453  &  0.476 & 0.426  &  0.347 & 0.336  &           \\
 			SDE                        &          & 0.507  &  0.348 & 0.452  &  0.457 & 0.387  &           \\
 			Min                        &          &-0.356  & -2.339 &-0.233  & -0.331 &-0.746  &           \\
 			Max                        &          & 2.053  &  0.214 & 1.954  &  1.889 & 1.747  &
 		\end{tabular}
 	\end{ruledtabular}
 	\fnt[1]{Values from Ref.~\onlinecite{marie2024val}.}
 	\fnt[2]{This work.}
 \end{table}
\end{squeezetable}

\subsection{Valence Ionization Potentials}

We first assess the performance of s$G_0W_0$ for the valence IP benchmark of Ref.~\onlinecite{marie2024val}, using the aug-cc-pVTZ basis set with HF and PBE0 starting points.
The resulting IPs are compared with the $G_0W_0$@HF and qs$GW$@HF results and FCI reference values reported in Ref.~\onlinecite{marie2024val}, as well as with $G_0W_0$@PBE0 results obtained in this work.

\begin{figure}
	\centering
	\includegraphics{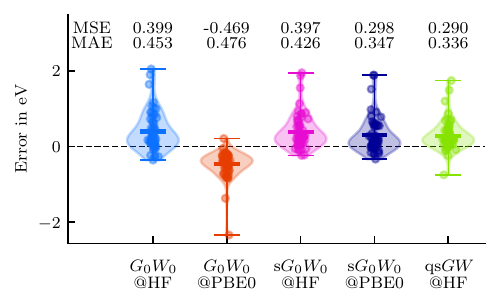}
  \caption{Violin plots of signed errors in valence IPs, relative to the FCI reference values reported in Ref.~\onlinecite{marie2024val}, for $G_0W_0$ and s$G_0W_0$ calculations using HF or PBE0 starting points, and for qs$GW$ starting from HF orbitals. All calculations use the aug-cc-pVTZ basis set.}
	\label{fig:violin_val}
	\vspace{1em}

\end{figure}

Table~\ref{tab:val_ip} lists the 58 valence IPs in order of increasing electron number, together with the summary error statistics relative to FCI. The corresponding distributions of signed errors are shown in Fig.~\ref{fig:violin_val}. Compared with $G_0W_0$@HF, s$G_0W_0$@HF yields a modest reduction in the mean absolute error (MAE), from \SI{0.453}{\eV} to \SI{0.426}{\eV}, and a more pronounced reduction in the standard deviation of the errors (SDE), from \SI{0.507}{\eV} to \SI{0.452}{\eV}. The mean signed error (MSE), by contrast, is essentially unchanged (\SI{0.399}{\eV} versus \SI{0.397}{\eV}). The minimum and maximum signed errors also move closer to zero, from $\SI{-0.356}{\eV}$ and \SI{2.053}{\eV} to $\SI{-0.233}{\eV}$ and \SI{1.954}{\eV}, respectively. On an individual-state basis, s$G_0W_0$@HF reduces the absolute error for 35 of the 58 IPs. Thus, the inclusion of singles produces a modest but consistent improvement in the aggregate error statistics, without a large change in the overall error distribution.

The qs$GW$@HF results remain more accurate on average, with an MAE of \SI{0.336}{\eV} and an SDE of \SI{0.387}{\eV}. Nevertheless, qs$GW$@HF exhibits a larger negative outlier ($-\SI{0.746}{\eV}$) than either one-shot HF-based approach.  
It is noteworthy that including singles in the ECC-based formulation does not deteriorate the average valence-IP accuracy, as corrections beyond $GW$ do not necessarily lead to systematic improvements. \cite{LewisBerkelbach2019}
The improvement in the average valence-IP accuracy is therefore modest, and distinct from the reduction in starting-point dependence discussed below. However, this agreement should not be interpreted as evidence that s$G_0W_0$ is equivalent to an orbital-optimized dRPA calculation: the singles amplitudes follow the approximate drCCSD equations rather than an explicit orbital-stationarity condition.

Finally, we assess the starting-point dependence by comparing HF- and PBE0-based calculations for each method.
For s$G_0W_0$, the MSE decreases from $\SI{0.397}{\eV}$ to $\SI{0.298}{\eV}$, a change of $\SI{-0.099}{\eV}$.
For $G_0W_0$, it changes from $\SI{0.399}{\eV}$ to $\SI{-0.469}{\eV}$, a change of $\SI{-0.868}{\eV}$ and a reversal in sign.
Because changes in MSE can conceal cancellations between individual-state shifts, we also compare the absolute HF--PBE0 differences, i.e. $|E_{\textrm{(s)}G_0W_0\textrm{@HF}} - E_{\textrm{(s)}G_0W_0\textrm{@PBE0}}|$, for each of the 58 IPs. Their mean decreases from \SI{0.868}{\eV} with $G_0W_0$ to \SI{0.112}{\eV} with s$G_0W_0$, a reduction of approximately 87\%. The largest absolute starting-point difference likewise decreases from \SI{2.566}{\eV} to \SI{0.459}{\eV}. This provides direct evidence that singles strongly suppress the starting-point sensitivity of individual valence IPs.

\subsection{Core Ionization Potentials}

We next consider the core-IP benchmark set of Ref.~\onlinecite{marie2026core}.
Both $G_0W_0$ and s$G_0W_0$ calculations were performed in the aug-cc-pCVTZ basis set using neutral, closed-shell HF and KS references.
The resulting IPs are compared in Table~\ref{tab:core_ip_1} with the core-valence-separated (CVS) IP-CC2, IP-CCSD, and IP-CC3 results and the CVS-FCI reference values of Ref.~\onlinecite{marie2026core}.
For symmetry-equivalent core orbitals, we used symmetry-adapted combinations rather than symmetry-broken orbitals.\cite{Cederbaum_1977a} As in Ref.~\onlinecite{marie2026core}, the tabulated IP corresponds to the higher of the two values obtained from the symmetric and antisymmetric solutions (i.e., the lower quasiparticle energy).

\begin{figure*}
	\centering
	\includegraphics{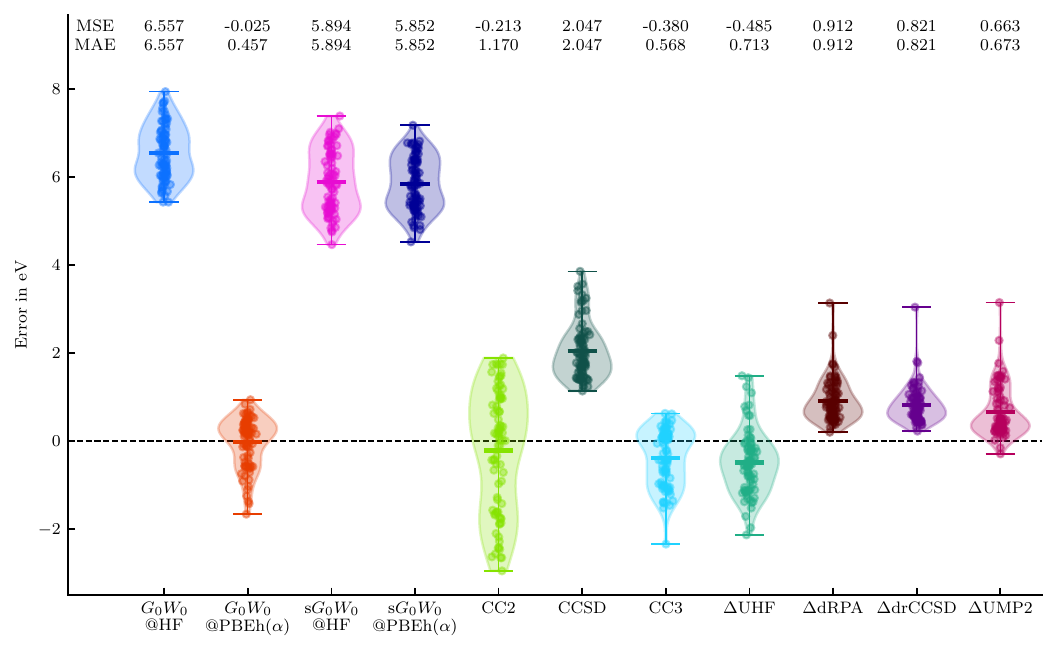}
  \caption{Violin plots of signed errors in core IPs, relative to the CVS-FCI reference values. The $G_0W_0$ and s$G_0W_0$ calculations use either HF or PBEh($\alpha$) as the starting point. The PBEh($\alpha$) starting points have been obtained using $\alpha = 0.45$ and $0.70$ for second- and third-row atoms, respectively. Results are also shown for CVS-IP-CC2, CVS-IP-CCSD, CVS-IP-CC3, $\Delta$UHF, 
    $\Delta$dRPA, $\Delta$drCCSD, and $\Delta$UMP2. The coupled-cluster reference values, $\Delta$UMP2 results, $G_0W_0$@PBEh($\alpha$) results, and CVS-FCI values are taken from Ref.~\onlinecite{marie2026core}. All calculations use the aug-cc-pCVTZ basis set.}
	\label{fig:violin_core}
\end{figure*}

The statistics reported in Table~\ref{tab:core_ip_1} and the corresponding error distributions in Fig.~\ref{fig:violin_core} use the tabulated CVS-FCI values as references. Unlike the CVS-IP-CC calculations, the $G_0W_0$ and s$G_0W_0$ calculations do not impose the CVS restriction, so the methods do not employ identical configuration spaces. Nevertheless, the comparison reveals a substantially larger singles effect for core ionizations than for valence ionizations. The MSE decreases from \SI{6.557}{\eV} for $G_0W_0$ to \SI{5.894}{\eV} for s$G_0W_0$, corresponding to an average downward shift of approximately \SI{0.663}{\eV}. The minimum and maximum signed errors decrease from \SI{5.433}{\eV} and \SI{7.944}{\eV} to \SI{4.467}{\eV} and \SI{7.390}{\eV}, respectively. Because all signed errors remain positive, the MSE and MAE are identical for these two methods. In contrast to the valence results, the SDE increases slightly, from 
\SI{0.611}{\eV} to \SI{0.703}{\eV}.

Thus, singles reduce the pronounced positive bias of $G_0W_0$ for core IPs, but they do not remove it: s$G_0W_0$ still overestimates every IP in this benchmark, and its absolute errors remain substantially larger than those of the CVS-IP-CC methods. The improvement should therefore be interpreted primarily as a systematic shift of the predicted core IPs rather than a uniform improvement in their spread.

We assess the starting-point dependence of $G_0W_0$ and s$G_0W_0$ by comparing the core IPs obtained with HF and PBEh($\alpha$) references.
For s$G_0W_0$, switching from HF to PBEh($\alpha$) reduces the MSE by $\SI{0.042}{\eV}$, from $\SI{5.894}{\eV}$ to $\SI{5.852}{\eV}$.
For $G_0W_0$, the corresponding reduction is $\SI{6.582}{\eV}$, from $\SI{6.557}{\eV}$ to $\SI{-0.025}{\eV}$.
More directly, the mean of the absolute differences $|E_{\textrm{(s)}G_0W_0\textrm{@HF}} - E_{\textrm{(s)}G_0W_0\textrm{@PBEh}(\alpha)}|$ for core IPs decreases from \SI{6.583}{\eV} for $G_0W_0$ to \SI{0.113}{\eV} for s$G_0W_0$, a reduction of approximately 98\%. The largest starting-point difference decreases from \SI{8.20}{\eV} to \SI{0.36}{\eV}. Hence, singles largely suppress the starting-point dependence at the level of individual core IPs, rather than merely reducing the mean signed shift.
This reduced sensitivity does not, however, translate into improved accuracy for a PBEh($\alpha$) starting point: the MAE increases from \SI{0.457}{\eV} for $G_0W_0$@PBEh($\alpha$) to \SI{5.852}{\eV} for s$G_0W_0$@PBEh($\alpha$), close to the s$G_0W_0$@HF result. The favorable performance of the tuned one-shot calculation must also be viewed in context: the fraction of exact exchange differs markedly between second- and third-row core ionizations (45\% and 70\%, respectively). Choosing this fraction without prior calibration or knowledge of accurate reference IPs is not straightforward. Such element-dependent tuning is therefore difficult to use as a predictive, transferable procedure, whereas s$G_0W_0$ largely removes the need to select a particular hybrid starting point.

\begin{squeezetable}
  \begin{table*}
    \setlength{\tabcolsep}{1pt}
    \caption{Core IPs (in \si{\eV}) calculated in the aug-cc-pCVTZ basis set. The $G_0W_0$ and s$G_0W_0$ calculations use either HF or PBEh($\alpha$) as the starting point and are compared with CVS-IP-CC2, CVS-IP-CCSD, CVS-IP-CC3, and CVS-FCI results from Ref.~\onlinecite{marie2026core}. An asterisk marks the core-ionized atom. Error statistics are calculated relative to CVS-FCI and reported in \si{\eV}: mean signed error (MSE), mean absolute error (MAE), standard deviation of the errors (SDE), and minimum and maximum signed errors (Min and Max). The PBEh($\alpha$) starting points have been obtained using $\alpha = 0.45$ and $0.70$ for second- and third-row atoms, respectively.}
  \label{tab:core_ip_1}
  	\begin{ruledtabular}
      \begin{tabular}{
        l|*{8}{S[table-format=4.2]}||
        l|*{8}{S[table-format=4.2]}
      }
               & \multicolumn{1}{c}{$G_0W_0$\rlap{\fnm[1]}} 
               & \multicolumn{1}{c}{$G_0W_0$\rlap{\fnm[2]\fnm[3]}}
               & \multicolumn{1}{c}{s$G_0W_0$\rlap{\fnm[1]}} 
               & \multicolumn{1}{c}{s$G_0W_0$\rlap{\fnm[1]}}    
               & \multicolumn{1}{c}{\multirow{2}{*}{CC2\rlap{\fnm[2]}}} 
               & \multicolumn{1}{c}{\multirow{2}{*}{CCSD\rlap{\fnm[2]}}} 
               & \multicolumn{1}{c}{\multirow{2}{*}{CC3\rlap{\fnm[2]}}} 
               & \multicolumn{1}{c||}{\multirow{2}{*}{FCI\rlap{\fnm[2]}}} 
               & 
               & \multicolumn{1}{c}{$G_0W_0$\rlap{\fnm[1]}} 
               & \multicolumn{1}{c}{$G_0W_0$\rlap{\fnm[2]}}
               & \multicolumn{1}{c}{s$G_0W_0$\rlap{\fnm[1]}} 
               & \multicolumn{1}{c}{s$G_0W_0$\rlap{\fnm[1]}}
               & \multicolumn{1}{c}{\multirow{2}{*}{CC2\rlap{\fnm[2]}}}
               & \multicolumn{1}{c}{\multirow{2}{*}{CCSD\rlap{\fnm[2]}}}
               & \multicolumn{1}{c}{\multirow{2}{*}{CC3\rlap{\fnm[2]}}} 
               & \multicolumn{1}{c}{\multirow{2}{*}{FCI\rlap{\fnm[2]}}} \\  
               & \multicolumn{1}{c}{@HF}
               & \multicolumn{1}{c}{@PBEh($\alpha$)} 
               & \multicolumn{1}{c}{@HF}
               & \multicolumn{1}{c}{@PBEh($\alpha$)}
               & & & & & 
               & \multicolumn{1}{c}{@HF}
               & \multicolumn{1}{c}{@PBEh($\alpha$)}
               & \multicolumn{1}{c}{@HF}
               & \multicolumn{1}{c}{@PBEh($\alpha$)}
               & & & & \\\hline 
        C$^*$H$_4$   & 296.43            & 290.96              & 295.69              & 295.75                & 291.64             & 291.99             & 290.80            & 290.85  &  CS$^*$        & 2478.72 & 2473.12 & 2478.36 & 2478.35  & 2472.82 & 2476.14 & 2472.14 & 2472.81  \\ 
        H$_2$O$^*$   & 546.22            & 538.99              & 545.76              & 545.66                & 537.98             & 541.48             & 538.87            & 539.87  &  C$^*$S        & 300.71  & 294.11  & 299.42  & 299.49   & 294.75  & 295.26  & 294.02  & 293.39   \\ 
        HF$^*$       & 700.10            & 692.44              & 699.91              & 699.73                & 691.15             & 695.81             & 692.74            & 694.10  &  H$_2$CCO$^*$  & 547.49  & 539.66  & 547.19  & 546.95   & 538.60  & 542.73  & 539.02  & 540.17   \\ 
        N$^*$H$_3$   & 411.84            & 405.30              & 411.15              & 411.14                & 405.10             & 407.04             & 405.17            & 405.63  &  H$_2$CC$^*$O  & 301.53  & 295.30  & 300.41  & 300.53   & 296.63  & 296.75  & 295.21  & 294.74   \\ 
        BF$^*$       & 701.98            & 694.01              & 701.91              & 701.58                & 692.80             & 697.78             & 694.02            & 695.43  &  H$_2$C$^*$CO  & 297.12  & 291.40  & 296.33  & 296.35   & 291.85  & 292.59  & 291.12  & 291.15   \\ 
        C$^*_2$H$_2$ & 297.32            & 291.49              & 296.48              & 296.52                & 291.95             & 292.69             & 291.46            & 291.28  &  H$_2$CN$^*$N  & 416.93  & 409.87  & 415.97  & 416.00   & 409.67  & 411.75  & 409.52  & 409.43   \\ 
        CO$^*$       & 548.89            & 541.76              & 548.57              & 548.38                & 540.29             & 544.26             & 541.54            & 542.47  &  H$_2$CNN$^*$  & 414.85  & 407.27  & 413.95  & 413.90   & 407.31  & 409.52  & 407.16  & 407.12   \\ 
        C$^*$O       & 301.69            & 296.14              & 300.73              & 300.79                & 297.73             & 297.66             & 296.57            & 296.26  &  H$_2$C$^*$NN  & 297.80  & 291.91  & 296.98  & 297.02   & 292.29  & 293.26  & 291.76  & 291.66   \\ 
        HCN$^*$      & 413.47            & 406.66              & 412.61              & 412.58                & 406.06             & 408.33             & 406.55            & 406.77  &  HNCO$^*$      & 547.47  & 539.77  & 547.06  & 546.84   & 538.46  & 542.63  & 539.06  & 540.14   \\ 
        HC$^*$N      & 299.04            & 293.54              & 298.19              & 298.25                & 294.22             & 294.63             & 293.68            & 293.40  &  HN$^*$CO      & 413.01  & 406.43  & 412.36  & 412.26   & 405.74  & 408.18  & 405.96  & 406.40   \\ 
        N$^*_2$      & 416.26            & 409.75              & 415.32              & 415.33                & 409.54             & 411.28             & 410.05            & 409.88  &  HNC$^*$O      & 302.63  & 296.50  & 301.36  & 301.51   & 297.70  & 297.92  & 296.41  & 295.94   \\ 
        C$^*_2$H$_4$ & 296.90            & 291.06              & 296.14              & 296.18                & 291.80             & 292.27             & 290.89            & 290.82  &  N$^*_2$CH$_2$ & 415.20  & 408.37  & 414.41  & 414.39   & 408.01  & 409.93  & 408.02  & 407.99   \\ 
        CH$_2$N$^*$H & 412.30            & 405.52              & 411.63              & 411.56                & 405.16             & 407.28             & 405.18            & 405.59  &  N$_2$C$^*$H$_2$  & 298.99  & 293.36  & 298.16  & 298.23   & 293.91  & 294.32  & 293.06  & 292.92   \\ 
        C$^*$H$_2$NH & 298.63            & 292.90              & 297.82              & 297.88                & 293.77             & 294.01             & 292.75            & 292.58  &  NNO$^*$       & 548.31  & 540.85  & 548.18  & 547.84   & 539.16  & 543.69  & 540.06  & 541.44   \\ 
        CH$_2$O$^*$  & 546.21            & 538.74              & 545.90              & 545.68                & 537.50             & 541.43             & 538.28            & 539.37  &  NN$^*$O       & 420.28  & 413.18  & 418.80  & 418.98   & 412.78  & 414.79  & 412.84  & 412.60   \\ 
        C$^*$H$_2$O  & 300.49            & 294.84              & 299.62              & 299.71                & 295.95             & 295.94             & 294.73            & 294.46  &  N$^*$NO       & 416.27  & 408.99  & 415.03  & 415.08   & 408.27  & 410.72  & 408.79  & 408.70   \\ 
        HNO$^*$      & 548.56            & 541.03              & 548.12              & 547.96                & 539.98             & 543.56             & 540.89            & 541.60  &  P$^*$N        & 2153.27 & 2147.63 & 2152.55 & 2152.60  & 2147.57 & 2150.62 & 2147.19 & 2147.47  \\ 
        HN$^*$O      & 416.89            & 410.02              & 415.97              & 416.00                & 410.17             & 411.57             & 409.96            & 409.84  &  PN$^*$        & 413.43  & 405.88  & 412.38  & 412.32   & 404.46  & 407.75  & 405.10  & 405.74   \\ 
        CH$_3$F$^*$  & 699.09            & 691.32              & 699.03              & 698.75                & 689.92             & 694.73             & 691.11            & 692.57  &  Si$^*$O       & 1850.73 & 1845.73 & 1850.05 & 1850.11  & 1845.20 & 1848.18 & 1845.09 & 1845.30  \\ 
        C$^*$H$_3$F  & 299.37            & 293.84              & 298.62              & 298.69                & 294.82             & 294.89             & 293.67            & 293.56  &  SiO$^*$       & 544.48  & 536.46  & 544.30  & 543.94   & 534.55  & 539.87  & 534.86  & 537.20   \\ 
        CH$_3$O$^*$H & 545.68            & 538.48              & 545.30              & 545.14                & 537.30             & 540.87             & 537.99            & 539.04  &  CH$_2$CHF$^*$ & 699.88  & 691.72  & 699.81  & 699.49   & 690.67  & 695.51  & 691.57  & 693.09   \\ 
        C$^*$H$_3$OH & 298.26            & 292.74              & 297.54              & 297.60                & 293.55             & 293.76             & 292.53            & 292.50  &  CH$_2$C$^*$HF & 299.71  & 293.73  & 298.89  & 298.95   & 294.72  & 295.01  & 293.61  & 293.44   \\ 
        F$^*_2$      & 703.74            & 695.62              & 703.38              & 703.23                & 694.31             & 698.58             & 695.57            & 696.41  &  C$^*$H$_2$CHF & 297.05  & 291.26  & 296.29  & 296.33   & 291.97  & 292.46  & 291.06  & 291.03   \\ 
        H$_2$S$^*$   & 2477.78           & 2472.05             & 2477.19             & 2477.27               & 2471.87            & 2474.94            & 2471.06           & 2471.69 &  CH$_2$S$^*$   & 2477.36 & 2471.60 & 2476.98 & 2476.99  & 2471.39 & 2474.68 & 2470.41 & 2471.11  \\ 
        HCl$^*$      & 2827.39           & 2821.42             & 2826.84             & 2826.91               & 2821.32            & 2824.61            & 2820.71           & 2821.42 &  C$^*$H$_2$S   & 299.25  & 292.92  & 298.12  & 298.22   & 293.48  & 294.07  & 292.56  & 292.28   \\ 
        HOF$^*$      & 701.14            & 693.26              & 701.17              & 700.91                & 691.71             & 696.48             & 692.96            & 694.17  &  CH$_3$CHO$^*$ & 545.57  & 537.92  & 545.33  & 545.10   & 536.87  & 540.83  & 537.45  & 538.48   \\ 
        HO$^*$F      & 549.97            & 542.41              & 549.27              & 549.21                & 541.64             & 544.49             & 542.07            & 542.55  &  CH$_3$C$^*$HO & 300.11  & 294.41  & 299.27  & 299.34   & 295.42  & 295.57  & 294.27  & 293.97   \\ 
        NH$_2$F$^*$  & 699.75            & 691.94              & 699.78              & 699.50                & 690.41             & 695.28             & 691.58            & 692.97  &  C$^*$H$_3$CHO & 297.27  & 291.76  & 296.52  & 296.55   & 292.39  & 292.76  & 291.44  & 291.36   \\ 
        N$^*$H$_2$F  & 415.29            & 408.62              & 414.50              & 414.51                & 408.74             & 410.13             & 408.34            & 408.44  &  CH$_3$NO$^*$  & 547.70  & 539.85  & 547.39  & 547.19   & 539.09  & 542.76  & 539.66  & 540.42   \\ 
        P$^*$H$_3$   & 2151.62           & 2146.36             & 2151.09             & 2151.16               & 2146.06            & 2148.92            & 2145.22           & 2145.73 &  CH$_3$N$^*$O  & 416.35  & 409.37  & 415.48  & 415.47   & 409.48  & 411.02  & 409.18  & 409.12   \\ 
        Si$^*$H$_4$  & 1849.13           & 1843.96             & 1848.68             & 1848.75               & 1843.46            & 1846.26            & 1842.87           & 1843.30 &  C$^*$H$_3$NO  & 298.12  & 292.45  & 297.29  & 297.34   & 293.17  & 293.48  & 292.10  & 292.02   \\ 
        CH$_3$CN$^*$ & 412.51            & 405.54              & 411.73              & 411.68                & 405.13             & 407.39             & 405.36            & 405.65  &  F$^*$CHO      & 700.96  & 692.84  & 700.77  & 700.47   & 691.51  & 696.44  & 692.54  & 693.94   \\ 
        CH$_3$C$^*$N & 298.50            & 293.04              & 297.74              & 297.77                & 293.46             & 294.03             & 292.99            & 292.79  &  FCHO$^*$      & 546.95  & 539.52  & 546.62  & 546.42   & 538.33  & 542.20  & 539.11  & 540.10   \\ 
        C$^*$H$_3$CN & 298.44            & 292.95              & 297.63              & 297.67                & 293.74             & 294.16             & 292.87            & 292.77  &  FC$^*$HO      & 303.28  & 297.52  & 302.26  & 302.38   & 298.77  & 298.72  & 297.42  & 297.03   \\ 
        CH$_3$N$^*$C & 412.96            & 406.65              & 412.44              & 412.33                & 405.74             & 408.26             & 406.28            & 406.71  &  H$_2$NCHO$^*$ & 544.71  & 536.96  & 544.48  & 544.23   & 535.91  & 540.02  & 536.43  & 537.56   \\ 
        C$^*$H$_3$NC & 299.04            & 293.59              & 298.36              & 298.39                & 294.33             & 294.71             & 293.42            & 293.36  &  H$_2$N$^*$CHO & 413.05  & 406.48  & 412.44  & 412.35   & 406.11  & 408.23  & 406.10  & 406.53   \\ 
        CH$_3$NC$^*$ & 298.56            & 292.53              & 297.65              & 297.68                & 293.88             & 294.07             & 292.67            & 292.33  &  H$_2$NC$^*$HO & 300.76  & 294.85  & 299.77  & 299.89   & 296.05  & 296.20  & 294.80  & 294.48   \\ 
        CO$^*_2$     & 548.31            & 540.85              & 547.85              & 547.65                & 539.38             & 543.46             & 540.20            & 541.24  &  HCOO$^*$H     & 547.59  & 540.20  & 547.17  & 546.96   & 539.07  & 542.70  & 539.71  & 540.68   \\ 
        C$^*$O$_2$   & 304.15            & 298.21              & 302.80              & 302.96                & 299.46             & 299.57             & 298.13            & 297.71  &  HCO$^*$OH     & 545.85  & 538.27  & 545.58  & 545.35   & 537.12  & 541.13  & 537.79  & 538.87   \\ \cline{1-9}
                     &                   &                     &                     &                       &                    &                    &                   &         &  HC$^*$OOH     & 302.01  & 296.21  & 301.02  & 301.14   & 297.41  & 297.46  & 296.12  & 295.68   \\ 
        MSE          & \statcell{6.557}  & \statcell{-0.025}   & \statcell{5.894}    & \statcell{5.852}      & \statcell{-0.213}  & \statcell{2.047}   & \statcell{-0.380} &         &  OCS$^*$       & 2478.39 & 2472.70 & 2477.91 & 2477.91  & 2472.28 & 2475.61 & 2471.43 & 2472.09  \\ 
        MAE          & \statcell{6.557}  & \statcell{ 0.457}   & \statcell{5.894}    & \statcell{5.852}      & \statcell{ 1.170}  & \statcell{2.047}   & \statcell{ 0.568} &         &  O$^*$CS       & 548.19  & 539.99  & 547.64  & 547.43   & 538.83  & 543.22  & 539.10  & 540.25   \\ 
        SDE          & \statcell{0.611}  & \statcell{ 0.564}   & \statcell{0.703}    & \statcell{0.601}      & \statcell{ 1.388}  & \statcell{0.628}   & \statcell{ 0.638} &         &  OC$^*$S       & 302.86  & 296.21  & 301.35  & 301.52   & 297.11  & 297.68  & 295.99  & 295.37   \\ 
        Min          & \statcell{5.433}  & \statcell{-1.660}   & \statcell{4.467}    & \statcell{4.528}      & \statcell{-2.950}  & \statcell{1.140}   & \statcell{-2.340} &         &  P$^*_2$       & 2152.32 & 2146.55 & 2151.55 & 2151.59  & 2146.10 & 2149.47 & 2145.03 & 2145.61  \\ 
        Max          & \statcell{7.944}  & \statcell{ 0.940}   & \statcell{7.390}    & \statcell{7.181}      & \statcell{ 1.890}  & \statcell{3.860}   & \statcell{ 0.630} &         &  Cl$^*_2$      & 2828.58 & 2822.45 & 2828.00 & 2828.03  & 2822.29 & 2825.69 & 2821.25 & 2822.27  
  		\end{tabular}                                          
  	\end{ruledtabular}                                       
  \fnt[1]{This work.}                                        
  \fnt[2]{Values from Ref.~\onlinecite{marie2026core}.}      
  \fnt[3]{Interchanging the core IPs assigned to C$^*$H$_3$NC and CH$_3$NC$^*$ in our analysis reproduces the $G_0W_0$@PBEh($\alpha$) error statistics reported in Ref.~\onlinecite{marie2026core}.}
  \end{table*}                                               
\end{squeezetable}                                           

In addition to the quasiparticle calculations, we evaluated core IPs from total-energy differences between the neutral and core-ionized states at the $\Delta$UHF, $\Delta$dRPA (equivalently, $\Delta$drCCD), and $\Delta$drCCSD levels. 
The $\Delta$UMP2 results of Ref.~\onlinecite{marie2026core} are included for comparison. 
The resulting IPs are listed in Table~\ref{tab:core_delta}.
The corresponding statistical errors are reported at the bottom of the table, and their distributions are included in Fig.~\ref{fig:violin_core}. For $\Delta$dRPA and $\Delta$drCCSD, the Si-core ionization of \ce{SiH4} is excluded from the statistical analysis because its energy differs anomalously from the other results. The associated doubles amplitudes have a norm exceeding unity, $\norm{T_2}>1$, suggesting that the computed solution may not be continuously connected to the intended reference state. 

The corresponding statistical errors are reported at the bottom of Table~\ref{tab:core_delta}, and their distributions are included in Fig.~\ref{fig:violin_core}. For $\Delta$dRPA and $\Delta$drCCSD, the Si-core ionization of \ce{SiH4} is excluded from the statistical analysis because its energy differs anomalously from the other results. The associated doubles amplitudes have a norm exceeding unity, $\norm{T_2}>1$, suggesting that the computed solution may not be continuously connected to the intended reference state.

Some core-ionized drCCD and drCCSD calculations also yield complex correlation energies. Such behavior is associated with instabilities of the underlying RPA problem and the appearance of complex eigenvalues.\cite{burth2026} 
In these cases, we report the real part of the corresponding energy, which is also included in the statistical analysis. 
For several systems, including \ce{CH3OH}, \ce{CO2}, \ce{H2CCO}, \ce{HCOOH}, and \ce{OCS}, the instability arises only after singles are included, whereas instabilities already present at the drCCD level persist in drCCSD. 
These are not conventional spin-flip triplet instabilities, since the direct approximation excludes the corresponding spin-flip coupling. 
As discussed in Ref.~\onlinecite{burth2026}, imposing CVS does not necessarily remove such instabilities.

After excluding the anomalous \ce{SiH4} point, $\Delta$dRPA overestimates the core IPs by \SI{0.912}{\eV} on average, while the inclusion of singles reduces the MSE and MAE to \SI{0.821}{\eV} at the $\Delta$drCCSD level. Both approaches have exclusively positive signed errors in the retained set. The SDE also decreases, from \SI{0.471}{\eV} for $\Delta$dRPA to \SI{0.406}{\eV} for $\Delta$drCCSD. Nevertheless, $\Delta$UHF yields a smaller MAE of \SI{0.713}{\eV}, although its SDE is larger (\SI{0.718}{\eV}). The $\Delta$UMP2 results are slightly more accurate still, with an MAE of \SI{0.673}{\eV} according to Table~\ref{tab:core_delta}. Consequently, the singles correction improves the direct-ring total-energy differences, but this improvement is not sufficient to outperform the simpler $\Delta$UHF and $\Delta$UMP2 approaches on average.

\begin{squeezetable}
  \begin{table*}
    \caption{Core IPs (in \si{\eV}) obtained as total-energy differences between the core-ionized and neutral states, computed in the aug-cc-pCVTZ basis set at the $\Delta$UHF, $\Delta$dRPA, $\Delta$drCCSD, and $\Delta$UMP2 levels. An asterisk identifies the core-ionized atom; superscripts refer to the notes below the table. The statistical analysis utilizing errors relative to CVS-FCI (in \si{\eV}) is characterized by the mean signed error (MSE), mean absolute error (MAE), standard deviation of the errors (SDE), and minimum and maximum signed errors (Min and Max). The $\Delta$UHF and $\Delta$UMP2 statistics include 84 ionizations; the $\Delta$dRPA and $\Delta$drCCSD statistics exclude the anomalous Si-core ionization of SiH$_4$ ($N=83$).}	\label{tab:core_delta}
  	\begin{ruledtabular}
  		\begin{tabular}{l|rrrrr||l|rrrrr}
        & \multicolumn{1}{c}{$\Delta$UHF\rlap{\fnm[1]}}
        & \multicolumn{1}{c}{$\Delta$dRPA\rlap{\fnm[1]}}
        & \multicolumn{1}{c}{$\Delta$drCCSD\rlap{\fnm[1]}}
        & \multicolumn{1}{c}{$\Delta$UMP2\rlap{\fnm[2]}}
        & \multicolumn{1}{c||}{CVS-FCI\fnm[2]}
        &
        & \multicolumn{1}{c}{$\Delta$UHF\rlap{\fnm[1]}}
        & \multicolumn{1}{c}{$\Delta$dRPA\rlap{\fnm[1]}}
        & \multicolumn{1}{c}{$\Delta$drCCSD\rlap{\fnm[1]}}
        & \multicolumn{1}{c}{$\Delta$UMP2\rlap{\fnm[2]}}
        & \multicolumn{1}{c}{CVS-FCI\fnm[2]}\\ \hline
        C$^*$H$_4$   & 290.61                       & 291.65                        & 291.64                        & 290.98  & 290.85 & CS$^*$          & 2472.57                       & 2473.40                        & 2473.40                            & 2473.36 & 2472.81 \\
        H$_2$O$^*$   & 539.01                       & 540.45                        & 540.44                        & 540.06  & 539.87 & C$^*$S          & 294.63                        & 294.89                         & 294.66                             & 293.74  & 293.39  \\
        HF$^*$       & 692.94                       & 694.60                        & 694.61                        & 694.34  & 694.10 & H$_2$CCO$^*$    & 538.99                        & 540.71                         & 540.62{\rlap{\fnm[4]}}             & 540.66  & 540.17  \\
        N$^*$H$_3$   & 405.12                       & 406.32                        & 406.31                        & 405.78  & 405.63 & H$_2$CC$^*$O    & 294.35                        & 296.48{\rlap{\fnm[4]}}         & 296.19{\rlap{\fnm[4]}}             & 296.18  & 294.74  \\
        BF$^*$       & 694.32                       & 695.97                        & 695.99                        & 695.86  & 695.43 & H$_2$C$^*$CO    & 290.91                        & 291.93                         & 291.91                             & 291.23  & 291.15  \\
        C$^*_2$H$_2$ & 290.72{\rlap{\fnm[3]}}       & 292.49{\rlap{\fnm[3]}}        & 292.26{\rlap{\fnm[3]}}        & 292.29  & 291.28 & H$_2$CN$^*$N    & 409.21                        & 410.51                         & 410.41                             & 410.40  & 409.43  \\
        CO$^*$       & 541.30                       & 542.97                        & 542.96                        & 542.92  & 542.47 & H$_2$CNN$^*$    & 406.45                        & 408.39                         & 408.15                             & 408.51  & 407.12  \\
        C$^*$O       & 296.40                       & 298.02                        & 297.60                        & 297.75  & 296.26 & H$_2$C$^*$NN    & 291.30                        & 292.61                         & 292.51                             & 292.18  & 291.66  \\
        HCN$^*$      & 406.07                       & 407.69                        & 407.52                        & 407.53  & 406.77 & HNCO$^*$        & 539.35                        & 540.78                         & 540.76                             & 540.50  & 540.14  \\
        HC$^*$N      & 292.88                       & 294.86                        & 294.55                        & 294.99  & 293.40 & HN$^*$CO        & 405.95                        & 407.11                         & 407.10                             & 406.58  & 406.40  \\
        N$^*_2$      & 409.42{\rlap{\fnm[3]}}       & 411.13{\rlap{\fnm[3]}}        & 410.86{\rlap{\fnm[3]}}        & 411.09  & 409.88 & HNC$^*$O        & 296.63{\rlap{\fnm[3]}}        & 298.34{\rlap{\fnm[3]}}         & 297.72{\rlap{\fnm[3]}}             & 298.23  & 295.94  \\
        C$^*_2$H$_4$ & 290.27{\rlap{\fnm[3]}}       & 291.64{\rlap{\fnm[3]}}        & 291.57{\rlap{\fnm[3]}}        & 291.23  & 290.82 & N$^*_2$CH$_2$   & 407.19{\rlap{\fnm[3]}}        & 409.04{\rlap{\fnm[3]}}         & 408.86{\rlap{\fnm[3]}}             & 409.12  & 407.99  \\
        CH$_2$N$^*$H & 404.71                       & 406.16                        & 406.13                        & 405.83  & 405.59 & N$_2$C$^*$H$_2$ & 292.74                        & 293.82                         & 293.78                             & 293.26  & 292.92  \\
        C$^*$H$_2$NH & 292.02                       & 293.70                        & 293.55                        & 293.45  & 292.58 & NNO$^*$         & 540.31{\rlap{\fnm[3]}}        & 542.29{\rlap{\fnm[3]}}         & 542.12{\rlap{\fnm[3]}}             & 542.56  & 541.44  \\
        CH$_2$O$^*$  & 537.99                       & 539.75                        & 539.75                        & 539.65  & 539.37 & NN$^*$O         & 413.70{\rlap{\fnm[5]}}        & 413.93{\rlap{\fnm[5]}}         & 413.80{\rlap{\fnm[5]}}             & 414.37  & 412.60  \\
        C$^*$H$_2$O  & 293.87                       & 295.78                        & 295.60                        & 295.68  & 294.46 & N$^*$NO         & 408.54                        & 410.07                         & 409.79                             & 410.06  & 408.70  \\
        HNO$^*$      & 540.29                       & 542.10                        & 542.07                        & 542.10  & 541.60 & P$^*$N          & 2148.05                       & 2148.07                        & 2148.03                            & 2147.18 & 2147.47  \\
        HN$^*$O      & 408.77{\rlap{\fnm[3]}}       & 411.06{\rlap{\fnm[3]}}        & 410.88{\rlap{\fnm[3]}}        & 411.29  & 409.84 & PN$^*$          & 404.61                        & 406.55                         & 406.35                             & 406.90  & 405.74  \\
        CH$_3$F$^*$  & 691.17                       & 692.99                        & 693.00                        & 692.86  & 692.57 & Si$^*$O         & 1846.12                       & 1848.44                        & 1848.35                            & 1845.81 & 1845.3  \\
        C$^*$H$_3$F  & 293.50                       & 294.46                        & 294.44                        & 293.84  & 293.56 & SiO$^*$         & 535.68                        & 537.40                         & 537.43                             & 537.52  & 537.20  \\
        CH$_3$O$^*$H & 537.99                       & 539.54                        & 539.54                        & 539.26  & 539.04 & CH$_2$CHF$^*$   & 691.71                        & 693.52                         & 693.52                             & 693.43  & 693.09  \\
        C$^*$H$_3$OH & 292.29                       & 293.33                        & 293.31{\rlap{\fnm[4]}}        & 292.74  & 292.50 & CH$_2$C$^*$HF   & 292.98                        & 294.35                         & 294.24                             & 294.02  & 293.44  \\
        F$^*_2$      & 694.28{\rlap{\fnm[3]\fnm[5]}}& 696.93{\rlap{\fnm[3]\fnm[5]}} & 696.82{\rlap{\fnm[3]\fnm[5]}} & 697.13  & 696.41 & C$^*$H$_2$CHF   & 290.60                        & 291.83                         & 291.78                             & 291.30  & 291.03  \\
        H$_2$S$^*$   & 2471.78                      & 2472.39                       & 2472.40                       & 2472.01 &2471.69 & CH$_2$S$^*$     & 2470.74                       & 2471.65                        & 2471.64                            & 2471.51 & 2471.11 \\
        HCl$^*$      & 2821.38                      & 2822.19                       & 2822.19                       & 2821.81 &2821.42 & C$^*$H$_2$S     & 291.70                        & 293.37                         & 293.23                             & 293.22  & 292.28  \\
        HOF$^*$      & 692.20                       & 694.49                        & 694.47                        & 694.68  & 694.17 & CH$_3$CHO$^*$   & 537.19                        & 538.96{\rlap{\fnm[4]}}         & 538.94{\rlap{\fnm[4]}}             & 538.84  & 538.48  \\
        HO$^*$F      & 541.32                       & 543.56                        & 543.39                        & 543.74  & 542.55 & CH$_3$C$^*$HO   & 293.44                        & 295.37                         & 295.17                             & 295.38  & 293.97  \\
        NH$_2$F$^*$  & 691.26                       & 693.33                        & 693.33                        & 693.36  & 692.97 & C$^*$H$_3$CHO   & 291.31                        & 292.42                         & 292.34                             & 291.81  & 291.36  \\
        N$^*$H$_2$F  & 408.03                       & 409.56                        & 409.40                        & 409.29  & 408.44 & CH$_3$NO$^*$    & 539.08                        & 540.90                         & 540.86                             & 540.90  & 540.42  \\
        P$^*$H$_3$   & 2146.00                      & 2146.40                       & 2146.40                       & 2146.05 &2145.73 & CH$_3$N$^*$O    & 409.13{\rlap{\fnm[5]}}        & 410.29{\rlap{\fnm[5]}}         & 410.11{\rlap{\fnm[5]}}             & 410.60  & 409.12  \\
        Si$^*$H$_4$  & 1843.56                      & 1827.47                       & 1826.35                       & 1844.00 &1843.30 & C$^*$H$_3$NO    & 292.03                        & 292.93                         & 292.91                             & 292.20  & 292.02  \\
        CH$_3$CN$^*$ & 404.91                       & 406.40                        & 406.28                        & 406.14  & 405.65 & F$^*$CHO        & 692.96                        & 694.62                         & 694.55                             & 694.40  & 693.94  \\
        CH$_3$C$^*$N & 292.65                       & 293.70                        & 293.63                        & 293.10  & 292.79 & FCHO$^*$        & 538.99                        & 540.59                         & 540.58                             & 540.39  & 540.10  \\
        C$^*$H$_3$CN & 292.21                       & 294.22                        & 293.89                        & 294.26  & 292.77 & FC$^*$HO        & 296.97{\rlap{\fnm[3]}}        & 298.35{\rlap{\fnm[3]\fnm[4]}}  & 298.84{\rlap{\fnm[3]\fnm[4]}}      & 298.52  & 297.03  \\
        CH$_3$N$^*$C & 405.79                       & 407.30                        & 407.27                        & 407.13  & 406.71 & H$_2$NCHO$^*$   & 536.43                        & 538.03                         & 538.03                             & 537.87  & 537.56  \\
        C$^*$H$_3$NC & 293.17                       & 294.21                        & 294.20                        & 293.88  & 293.36 & H$_2$N$^*$CHO   & 406.04                        & 407.26                         & 407.23                             & 406.77  & 406.53  \\
        CH$_3$NC$^*$ & 292.10                       & 294.02                        & 293.66                        & 293.68  & 292.33 & H$_2$NC$^*$HO   & 295.07                        & 295.56                         & 295.54                             & 294.75  & 294.48  \\
        CO$^*_2$     & 540.53{\rlap{\fnm[3]}}       & 541.85{\rlap{\fnm[3]}}        & 541.84{\rlap{\fnm[3]\fnm[4]}} & 541.47  & 541.24 & HCOO$^*$H       & 539.91                        & 541.31                         & 541.28                             & 540.95  & 540.68  \\
        C$^*$O$_2$   & 299.15                       & 299.05                        & 298.96                        & 297.56  & 297.71 & HCO$^*$OH       & 537.76                        & 539.34                         & 539.33                             & 539.15  & 538.87  \\ \cline{1-6}
                     &                              &                               &                               &         &        & HC$^*$OOH       & 296.47                        & 297.16                         & 297.02{\rlap{\fnm[4]}}             & 296.49  & 295.68  \\
        MSE          &   -0.485                     & 0.912                         & 0.821                         &  0.663  &        & OCS$^*$         & 2472.16                       & 2472.75                        & 2472.76                            & 2472.40 & 2472.09 \\
        MAE          &    0.713                     & 0.912                         & 0.821                         &  0.673  &        & O$^*$CS         & 539.45{\rlap{\fnm[3]}}        & 541.11{\rlap{\fnm[3]}}         & 540.96{\rlap{\fnm[3]\fnm[4]}}      & 541.17  & 540.25  \\
        SDE          &    0.718                     & 0.471                         & 0.406                         &  0.570  &        & OC$^*$S         & 296.86                        & 296.82                         & 296.67                             & 295.39  & 295.37  \\
        Min          &   -2.131                     & 0.203                         & 0.228                         & -0.290  &        & P$^*_2$         & 2145.46{\rlap{\fnm[3]\fnm[5]}}& 2146.37{\rlap{\fnm[3]\fnm[5]}} & 2146.19{\rlap{\fnm[3]\fnm[5]}}     & 2148.76 & 2145.61 \\
        Max          &    1.485                     & 3.139                         & 3.045                         &  3.150  &        & Cl$^*_2$        & 2821.81{\rlap{\fnm[3]}}       & 2822.71{\rlap{\fnm[3]}}        & 2822.71{\rlap{\fnm[3]}}            & 2822.49 & 2822.27  
  		\end{tabular}
  	\end{ruledtabular}
    \fnt[1]{This work.}
    \fnt[2]{Values from Ref.~\onlinecite{marie2026core}.}
  	\fnt[3]{UHF calculation for the core-ionized state performed using QuAcK.\cite{QuAcK}}
  	\fnt[4]{Imaginary contribution to the energy arising from an instability of the RPA matrix.}
  	\fnt[5]{Reference energy of the core-ionized state reported in Ref.~\onlinecite{marie2026core} could not be reproduced.}
  \end{table*}
\end{squeezetable}

\section{Conclusion}
\label{sec:conclusion}

In this work, we have introduced single excitations into the extended coupled-cluster representation of $G_0W_0$ to incorporate correlation-driven orbital relaxation without iterating the Green's function to self-consistency.
Starting from the IP/EA-EOM-drECCD formulation, we have included right-hand single-excitation and left-hand single-deexcitation cluster amplitudes in the ground-state parametrization while retaining the direct-ring treatment of the doubles amplitudes.
The resulting drCCSD doubles-amplitude equations admit a generalized RPA-like formulation, and the corresponding IP/EA-EOM construction yields the singles-$G_0W_0$ (s$G_0W_0$) approximation.

We have assessed this approach for molecular valence and core IPs using HF and KS starting points.
For the 58 valence IPs, s$G_0W_0$ provides a modest improvement over $G_0W_0$, reducing the MAE from \SI{0.453}{\eV} to \SI{0.426}{\eV}, although qs$GW$ remains more accurate on average.
More importantly, the inclusion of singles substantially reduces the dependence on the starting point, bringing the s$G_0W_0$@PBE0 quasiparticle energies close to their s$G_0W_0$@HF counterparts.
The effect of singles is more pronounced for the 84 core IPs, for which the MAE decreases from \SI{6.557}{\eV} to \SI{5.894}{\eV}.
Nevertheless, both methods systematically overestimate the core IPs, and the spread of the errors increases slightly upon including singles.
In complementary total-energy-difference calculations, the inclusion of singles also improves the direct-ring results for core ionizations, although the simpler $\Delta$UHF and $\Delta$UMP2 approaches remain more accurate on average.
Some core-ionized references exhibit instabilities and complex correlation energies, highlighting a further limitation of the present approximation.

Overall, these results demonstrate that singles can have
a significant effect on $GW$-like quasiparticle energies,
particularly for core ionizations, but do not by themselves
establish a general improvement in accuracy.
Instead, the most notable finding is that the
inclusion of singles largely suppresses the starting-point
dependence of $G_0W_0$ for both valence and core ionizations.
Regardless of whether an HF or KS
reference is employed, the resulting quasiparticle energies
remain close to those obtained from the HF starting point.
This behavior demonstrates that singles offer an effective
alternative to self-consistency for reducing the sensitivity
of $GW$-like approximations to the underlying mean-field
description, although this does not necessarily translate
into improved absolute accuracy.
In particular, the strong performance of tuned $G_0W_0$@PBEh($\alpha$) for core IPs relies on substantially different, element-dependent fractions of exact exchange. The present reduction in starting-point dependence is therefore especially useful from the standpoint of avoiding such a priori choices, even though the absolute core-IP errors remain significant.
Further work should also examine the stability of the
drCCSD equations and clarify the relationship between
the present singles treatment and orbital-optimized or
self-consistent $GW$ approaches.

\acknowledgments{
The authors would like to thank Johannes T\"olle for stimulating discussions.
This study has been partially supported through the grant NanoX No.~ANR-17-EURE-0009 in the framework of the ``Programme des Investissements d’Avenir''.
This work used the HPC resources from CALMIP (Toulouse) under allocation 2026-18005. 
}

\section*{Data availability statement}
The data that support the findings of this study are available within the article.

\clearpage
\appendix
\section{Direct two-electron intermediates}
\label{app:Wd_pqrs}

In this Appendix, we provide the explicit expressions
for the direct two-electron intermediates $\wWd_{pqrs}$
obtained from the $T_1$ similarity transformation.
These intermediates are expressed in terms of the
original two-electron integrals and the singles
amplitudes $t_i^a$.

To account for the different prefactors entering the
same-spin and mixed-spin contributions, we introduce
the spin-dependent factor
\begin{equation}
  \eta_{pq}
  =
  \frac{1}{1+\delta_{\sigma_p\sigma_q}}
  =
  \begin{cases}
    \frac12, & \sigma_p = \sigma_q \\
    1, & \sigma_p \neq \sigma_q
  \end{cases}
\end{equation}
where $\sigma_p$ denotes the spin of orbital $p$.
This factor appears only in the terms involving
the spin-dependent averaging prescription introduced
in Sec.~\ref{sec:orbital_relaxation}.

The working equations read
\begin{subequations}
\begin{align}
  \wWd_{ijab}
  &= \ERI{ij}{ab}
  \\
  \wWd_{ijka}
  &= \ERI{ij}{ka}
  + \sum_c \ERI{ij}{ca} t_k^c
  \\
  \wWd_{ciab}
  &= \ERI{ci}{ab}
  - \sum_k \ERI{ki}{ab} t_k^c
  \\
  \wWd_{abcd}
  &= \ERI{ab}{cd}
  - \sum_k
  \qty(
    \ERI{ak}{cd} t_k^b
    + \ERI{bk}{dc} t_k^a
  )
  + \sum_{kl} \ERI{kl}{cd} t_k^a t_l^b
  \\
  \wWd_{ijkl}
  &= \ERI{ij}{kl}
  + \sum_c
  \qty(
    \ERI{ij}{kc} t_l^c
    + \ERI{ji}{lc} t_k^c
  )
  + \sum_{cd} \ERI{ij}{cd} t_k^c t_l^d
  \\
  \wWd_{iabj}
  &= \ERI{ia}{bj}
  + \sum_c \ERI{ai}{cb} t_j^c
  - \sum_k \ERI{ki}{jb} t_k^a
  - \sum_{ck} \ERI{ik}{bc} t_k^a t_j^c
  \\
  \wWd_{aibj}
  &= \ERI{ai}{bj}
  + \sum_c \ERI{ai}{bc} t_j^c
  - \sum_k \ERI{ik}{jb} t_k^a
  - \sum_{ck} \ERI{ki}{bc} t_k^a t_j^c
  \\
  \wWd_{abij}
  &= \ERI{ab}{ij}
  - P^+_{ai,bj} \sum_k I^\mathrm{I}_{kaji} t_k^b
  + P^+_{ai,bj}
  \sum_c I^\mathrm{II}_{baci} t_j^c
  \\
  \wWd_{abci}
  &= \ERI{ab}{ci}
  + \sum_d \ERI{ab}{cd} t_i^d
  + \eta_{ab} P^-_{ab}
  \sum_k \aERI{ka}{ci} t_k^b
  - \sum_{kl} \ERI{kl}{ic} t_k^a t_l^b
  \nonumber\\
  &+ \eta_{ab} P^-_{ab}
  \sum_{dk} \aERI{bk}{cd} t_k^a t_i^d
  + \sum_k I^\mathrm{VI}_{kbci} t_k^a
  \\
  \wWd_{kaij}
  &= \ERI{ka}{ij}
  + \sum_l \ERI{lk}{ij} t_l^a
  - \eta_{ij} P^-_{ij}
  \sum_c \aERI{ka}{ci} t_j^c
  - \sum_{cd} \ERI{ak}{cd} t_i^c t_j^d
  \nonumber\\
  &+ \eta_{ij} P^-_{ij}
  \sum_{cl} \aERI{kl}{jc} t_i^c t_l^a
  - \sum_c I^\mathrm{VI}_{kacj} t_i^c
\end{align}
\end{subequations}

For compactness, we have introduced the following
auxiliary intermediates:
\begin{subequations}
\begin{align}
  I^\mathrm{I}_{kaij}
  &= \ERI{ka}{ij}
  + \frac12 \sum_l \ERI{lk}{ij} t_l^a
  + \eta_{ij} P^-_{ij}
  \sum_c
  \qty(
    \aERI{ka}{cj} t_i^c
    + I^\mathrm{III}_{kaci} t_j^c
  )
  \\
  I^\mathrm{II}_{abci}
  &= \ERI{ab}{ci}
  + \frac12 \sum_d \ERI{ab}{cd} t_i^d
  + \eta_{ab} P^-_{ab}
  \sum_k I^\mathrm{IV}_{kaci} t_k^b
  \\
  I^\mathrm{III}_{iabj}
  &= \sum_c \ERI{ai}{bc} t_j^c
  - \frac12 \sum_k I^\mathrm{V}_{ikjb} t_k^a
  \\
  I^\mathrm{IV}_{iabj}
  &= \sum_k \ERI{ki}{jb} t_k^a
  \\
  I^\mathrm{V}_{ijka}
  &= \sum_c \ERI{ij}{ca} t_k^c
  \\
  I^\mathrm{VI}_{iabj}
  &= \sum_{ck} \aERI{ik}{bc} t_j^c t_k^a
\end{align}
\end{subequations}

\section{Excitation/deexcitation symmetry of the singles-RPA matrix}
\label{app:sym_sRPA}
In this Appendix, we demonstrate that the eigenvalues
of the singles-RPA matrix occur in opposite-sign pairs,
such that if $\Omega_\mu$ is an eigenvalue, so is
$-\Omega_\mu$.

Consider the general matrix
\begin{equation}
  \bH =
  \begin{pmatrix}
    \bA & \bB \\
    -\bC & -\bA^\text{T}
  \end{pmatrix}
\end{equation}
where $\bB = \bB^\text{T}$ and $\bC = \bC^\text{T}$.
Introducing the nonsingular matrix
\begin{equation}
  \bK =
  \begin{pmatrix}
    \bO & \bI \\
    -\bI & \bO
  \end{pmatrix}
\end{equation}
where $\bI$ and $\bO$ denote identity and zero
matrices of the appropriate dimensions, respectively,
one readily verifies that
\begin{equation}
  \bH^\text{T}\bK + \bK\bH = \bO
\end{equation}
Consequently,
\begin{equation}
  \bH = -\bK^{-1}\bH^\text{T}\bK
\end{equation}
Since similarity transformations and transposition
preserve the eigenvalues, $\bH$ and $-\bH$
have identical spectra.
Therefore, the eigenvalues of the singles-RPA matrix
occur in pairs $(\Omega_\mu,-\Omega_\mu)$.

\section{One- and two-body intermediates including singles}
\label{app:Vd_pqrs}

In this Appendix, we provide the explicit expressions
for the one- and two-body intermediates
$\wFF_{pq}$ and $\wVd_{pqrs}$ entering the
s$G_0W_0$ quasiparticle equations.
These intermediates arise from the
$(1+\hLam_1)\widetilde{H}_1$ contribution to the
transformed Hamiltonian and are expressed in terms
of the $T_1$-transformed intermediates introduced
in Appendix~\ref{app:Wd_pqrs} and the left-hand
singles amplitudes $\lambda_a^i$.

The one-body intermediates read
\begin{subequations}
\begin{align}
  \wFF_{ai}
  &= \wF_{ai}
  + \sum_{ck} \wW_{acik} \lambda_c^k
  \\
  \wFF_{ia}
  &= \wF_{ia}
  + \sum_c \wF_{ca} \lambda_c^i
  - \sum_k \wF_{ik} \lambda_a^k
  + \sum_{ck} \wW_{icak} \lambda_c^k
  \\
  \wFF_{ij}
  &= \wF_{ij}
  + \sum_{ck} \wW_{jcik} \lambda_c^k
  + \sum_c \wF_{cj} \lambda_c^i
  \\
  \wFF_{ab}
  &= \wF_{ab}
  + \sum_{ck} \wW_{bcak} \lambda_c^k
  - \sum_k \wF_{ak} \lambda_b^k
\end{align}
\end{subequations}

Similarly, the direct two-body intermediates are
\begin{subequations}
\begin{align}
  \wVd_{abij}
  &= \wWd_{abij}
  \\
  \wVd_{kaij}
  &= \wWd_{kaij}
  + \sum_c \wWd_{caij} \lambda_c^k
  \\
  \wVd_{abci}
  &= \wWd_{abci}
  - \sum_k \wWd_{abki} \lambda_c^k
  \\
  \wVd_{ijka}
  &= \wWd_{ijka}
  + \sum_l \wWd_{ijlk} \lambda_a^l
  - \eta_{ij} P^-_{ij}
  \sum_c \wW_{icak} \lambda_c^j
  \\
  \wVd_{ciab}
  &= \wWd_{ciab}
  + \sum_d \wWd_{cdab} \lambda_d^i
  + \eta_{ab} P^-_{ab}
  \sum_k \wW_{icak} \lambda_b^k
  \\
  \wVd_{iabj}
  &= \wWd_{iabj}
  + \sum_c \wWd_{cabj} \lambda_c^i
  - \sum_k \wWd_{iakj} \lambda_b^k
  \\
  \wVd_{ijab}
  &= \ERI{ij}{ab}
  - P^+_{ai,bj}
  \sum_k \wWd_{ijkb} \lambda_a^k
  + P^+_{ai,bj}
  \sum_c \wWd_{cjab} \lambda_c^i
\end{align}
\end{subequations}

As in Appendix~\ref{app:Wd_pqrs}, the factors
$\eta_{ij}$ and $\eta_{ab}$ account for the
different same-spin and mixed-spin contributions.
They are equal to $1/2$ for same-spin pairs and
unity for mixed-spin pairs.

\section*{References}

\bibliography{singles}

\end{document}